\documentclass[structabstract]{aa}
\usepackage{graphicx}
\usepackage{txfonts}
\usepackage{dcolumn}
\usepackage{natbib}
\begin{document}

\title{Ks- and Lp-band polarimetry on stellar and bow-shock sources in the Galactic center}
\titlerunning{Stellar and bow-shock polarimetry in the GC}
\author{R. M. Buchholz \inst{1}
       \and 
       G. Witzel \inst{1}
       \and
       R. Sch\"odel \inst{2,1}
       \and 
       A. Eckart \inst{1,3}
       }

\institute{I. Physikalisches Institut, Universit\"at zu K\"oln,
           Z\"ulpicher Str. 77, 50937 K\"oln, Germany\\
	   \email{buchholz,eckart,witzel@ph1.uni-koeln.de}
	   \and Instituto de Astrof\'isica de Andaluc\'ia (CSIC), Glorieta de la Astronom\'ia s/n, E-18008 Granada, Spain\\
	   \email{rainer@iaa.es}
	   \and
           Max-Planck-Institut f\"ur Radioastronomie, 
           Auf dem H\"ugel 69, 53121 Bonn, Germany
	   }

\date{Received 05/09/2012, accepted 10/06/2013}
\abstract{Infrared observations of the Galactic center (GC) provide a unique opportunity to study stellar and bow-shock polarization effects
in a dusty environment.}
{The goals of this work are to present new Ks- and Lp-band polarimetry on an unprecedented number of sources in the central parsec of the
GC, thereby expanding our previous results in the H- and Ks-bands.}    
{We use AO-assisted Ks- and Lp-band observations, obtained at the ESO VLT. High precision photometry and the new polarimetric calibration
method for NACO allow us to map the polarization in a region of 8''$\times$25'' (Ks) resp. 26''$\times$28'' (Lp). These are the first polarimetric
observations of the GC in the Lp-band in 30 years, with vastly improved spatial resolution compared to previous results. This allows
resolved polarimetry on bright bow-shock sources in this area for the first time at this wavelength.}
{We find foreground polarization to be largely parallel to the Galactic plane (Ks-band: 6.1\% at 20$^{\circ}$, Lp-band: 4.5\% at
20$^{\circ}$), in good agreement with our previous findings and with older results. The previously described Lp-band excess in the
foregound polarization towards the GC could be confirmed here for a much larger number of sources. The bow-shock sources contained
in the FOV seem to show a different relation between the polarization in the observed wavelength bands than what was determined for the
foreground. This points to the different relevant polarization mechanisms. The resolved polarization patterns of IRS~5 and 10W match
the findings we presented earlier for IRS~1W. Additionally, intrinsic Lp-band polarization was measured for IRS~1W and 21, as well
as for other, less prominent MIR-excess sources (IRS~2S, 2L, 5NE). The new data offer support for the presumed bow-shock nature of
several of these sources (1W, 5, 5NE, 10W, 21) and for the model of bow-shock polarization presented in our last work.}{}


\maketitle
\section{Introduction}
At a distance of $\sim8.0$ kpc \citep{ghez2008,gillessen2009}, the center of the Milky Way is by far the closest galactic
nucleus. In its central parsec, a nuclear stellar cluster (NSC) has been found with a $\sim4.0 \times 10^6$ M$_{\sun}$
super-massive black hole at its dynamical center \citep{eckart2002,schoedel2002,schoedel2003,ghez2003,ghez2008,gillessen2009}.
This cluster exhibits similar properties to the NSCs found at the dynamical and photometric centers of other galaxies \citep{boeker2010,schoedel2010c}.\\ 
Near-infrared (NIR) polarimetric observations of this region have been conducted over the past 35 years (see \cite{buchholz2011} for an overview).
In general, the shorter wavelength bands (H and K) seem to be dominated by line-of-sight (LOS) effects, while localized effects become more
important with longer wavelengths \citep[see][]{capps1976,knacke1977,kobayashi1980,lebofsky1982}.
These first studies have already shown that the Galactic center (GC) is well suited to studying both interstellar polarization and intrinsically
polarized sources.\\ 
More than a decade later, higher resolution observations (0.25'') enabled \cite{eckart1995} to increase the number of polarimetrically observed GC sources dramatically (160 sources in the central 13''$\times$13'', only Ks-band). In a follow-up study, \cite{ott1999}
examined $\sim$ 40 bright sources in the central 20''$\times$20'' at 0.5'' resolution. Both studies confirmed a largely uniform
foreground polarization, but a more complex picture started to emerge as well: individual sources showed different polarization
parameters, such as a significantly higher polarization degree (IRS~21).\\
These results illustrated that intrinsic polarization is not limited to longer wavelengths, but that it also plays a role in the Ks-band.
Specifically, sources embedded in the Northern Arm and other bow-shock sources show signs of intrinsic polarization, with IRS~21
showing the strongest total Ks-band polarization of a bright GC source detected to date \citep[$\sim$ 10-16\% at 16$^{\circ}$]{eckart1995,ott1999}.\\
Using observations at even higher resolution, \cite{buchholz2011} presented H- and Ks-band polarimetry on 163 (H) respectively
194 (Ks) sources in the central 3''$\times$19'' (with an additional FOV in the Ks-band that was rotated by
45$^{\circ}$ containing 186 sources, see Fig.\ref{FigFOVwoll}). That study indicated a possible additional large scale local
contribution to the
polarization, possibly caused by emission from aligned dust grains in the Northern Arm of the Minispiral. The new data we present
here extend
this study to the Lp-band, while also expanding the Ks-band FOV. Our new study contains the first polarimetric Lp-band observations
of the GC in the last 30 years \citep[with the last data at that wavelength taken by][]{lebofsky1982}.\\
In general, three mechanisms can lead to NIR polarization in GC sources: emission by heated, non-spherical dust grains, scattering
(on spherical and/or aligned non-spherical grains), and dichroic extinction by aligned dust grains. The observed LOS polarization
can be explained by the third effect, while the first two cases can be regarded as intrinsic to the source, thereby allowing
conclusions about the source itself and its immediate environment. If a source is enclosed in an optically thick dust shell, dichroic
extinction can become important on a local scale as well \citep[see e.g.][]{whitney2002}.\\
Even 60 years after \cite{davis_greenstein1951} suggested paramagnetic dissipation as the basic mechanism that could cause the
observed, large scale grain alignment, the issue is still under discussion. In this process, the angular momentum of rapidly
spinning grains is aligned with the magnetic field. This necessitates a sufficient grain magnetization and angular momentum, and
it is still not clear what mechanisms can produce this \citep[see e.g.][for an overview of the proposed
processes that are expected to be relevant in different environments]{purcell1971,lazarian2003,lazarian2007}. Therefore, it remains 
difficult to reach exact conclusions for dust parameters and magnetic field strength, but at least the magnetic field orientation can
be determined. If the parameters change along the LOS, this further complicates the issue.\\
Using the polarization angle as a probe for the magnetic field orientation, \cite{nishiyama2009,nishiyama2010} mapped magnetic
fields in the innermost 20' resp. 2$^{\circ}$ of  the GC. These studies did not cover the central parsec due to insufficient
resolution.\\
The GC cannot be observed at visual wavelengths due to the strong extinction caused by dust grains on the LOS
(up to $A_V=40-50$ mag, depending on the assumed extinction law). This effects extends into the NIR as well, with a value around
3 mag in the Ks-band \citep[e.g.][]{scoville2003,schoedel2010b,fritz2011}. While the contribution to the total LOS extinction does
not depend strongly on the geometry of individual grains, only elongated and aligned grains can contribute to the observed LOS 
polarization. This already implies that there is most likely no simple linear increase of polarization degree with rising extinction,
especially since grain alignment does not have to be the same in every dust cloud along the LOS. \cite{heiles1987} and \cite{jones1989}
proposed a model where grain alignment is influenced by a constant and a random component of the galactic 
magnetic field. The polarization caused by the former would ''pile up'', while the contributions from the latter would cancel each
other out  along the LOS. This would lead to a decline of polarization efficiency (the relation between polarization degree and 
extinction) with rising extinction. \cite{jones1989} found that this takes the form of a power law relation with an exponent of
$\beta = -0.25$. In our recent study, we were able to confirm this relation for the sources in the GC \citep[see][]{buchholz2011}.
In that work as well as in our new study, we used a Ks-band extinction map of the central parsec recently presented by
\cite{schoedel2010b}.\\
The wavelength dependency of the polarization degree is also complex in general, but in the NIR, a power law relation seems to apply
\begin{table*}[!t]
\caption{\small Details of the observations used for this work. {\em Mode} denotes the wavelength
filter (H/Ks/Lp band, with central wavelength in $\mu$m) and the type of observations
(Wollaston polarimetry (W or Wr), the latter with a rotated FOV).
{\em N} is the number of exposures
that were taken with a given detector integration time ({\em DIT}). {\em NDIT} (in sec) denotes the number of
integrations that were averaged online by the read-out electronics during the observation. {\em CAM} indicates the
camera optics used (S13 or L27). 
  The Strehl ratio (if given) was measured using the {\em strehl} algorithm
  of the ESO eclipse package \citep[][publicly available at {\em http://www.eso.org/projects/aot/eclipse/distrib/index.html}]{devillard1997}, given here is the average value
  over all images in each dataset.}
\label{TabObservations}
\centering       
\begin{tabular}{l l l l r r r l r}
\hline\hline
& program & date & mode & N & NDIT & DIT & CAM & Strehl\\
\hline
H1 & 073.B-0084(A) & 2004-06-12 & H, W & 30 & 1 & 30 & S13 & 0.17\\
K1 & 083.B-0031(A) & 2009-05-18 & Ks, W & 143 & 4 & 10 & S13 & 0.27\\
K2 & 179.B-0261(A) & 2007-04-03 & Ks, Wr & 70 & 2 & 15 & S13 & 0.22\\
K3 & 086.C-0049(A) & 2011-03-19 & Ks, W & 37 & 2 & 15 & S13 & \\
L1 & 086.C-0049(A) & 2011-03-17 & Lp, W & 26 & 150 & 0.2 & L27 & \\
\end{tabular}
\end{table*}
\citep[see][]{martin1990}. That study also showed that the semi-empirical law applicable to polarization in the optical domain
\citep{serkowski1975,mathis1986} is a poor approximation in the NIR. At longer wavelengths, the power law relation also seems to fail,
with Lp-band polarization degrees exceeding the predicted values significantly. This effect even seems to increase towards the GC
\citep{jones1990}. The exact cause for this effect is still unknown, partly due to the lack of polarimetric observations in the L/M-band
in the last decades.\\
In \cite{buchholz2011}, we presented the first resolved polarimetric study of two extended and intrinsically polarized sources in the
GC, IRS~1W and 21. \cite{tanner2002,tanner2005} described IRS~21 as a bow-shock most likely created by a mass-losing Wolf-Rayet star. Separating
the observed polarization into a foreground and a source intrinsic component, we were able to show that emission from aligned grains
is probably the dominant mechanism that produces intrinsic polarization in both of these bow-shock sources.\\    
The aims of this work are to present the first $\sim$0.12'' resolution Lp-band polarization measurements of GC sources, as well as Ks-band
polarimetry on a large number of sources north of the FOV examined in our previous study. This allows us to study two additional sources
embedded in the Northern Arm of the Minispiral (IRS~5 and 10W), as well as other extended and intrinsically polarized objects. Resolved Lp-band polarimetry could be obtained on all bright bow-shock sources in the central 26''$\times$28''.\\
In \S \ref{SectObs}, we present the photometric and calibration methods applied to the data. The results are presented in
\S \ref{SectResults}, followed by a summary and discussion of their implications in \S \ref{SectDiscussion}.
\section{Observation and data reduction}
\label{SectObs}
\subsection{Observation}
\label{SectObservations}
\begin{figure*}[!t]
\centering
\includegraphics[width=\textwidth,angle=0, scale=1.0]{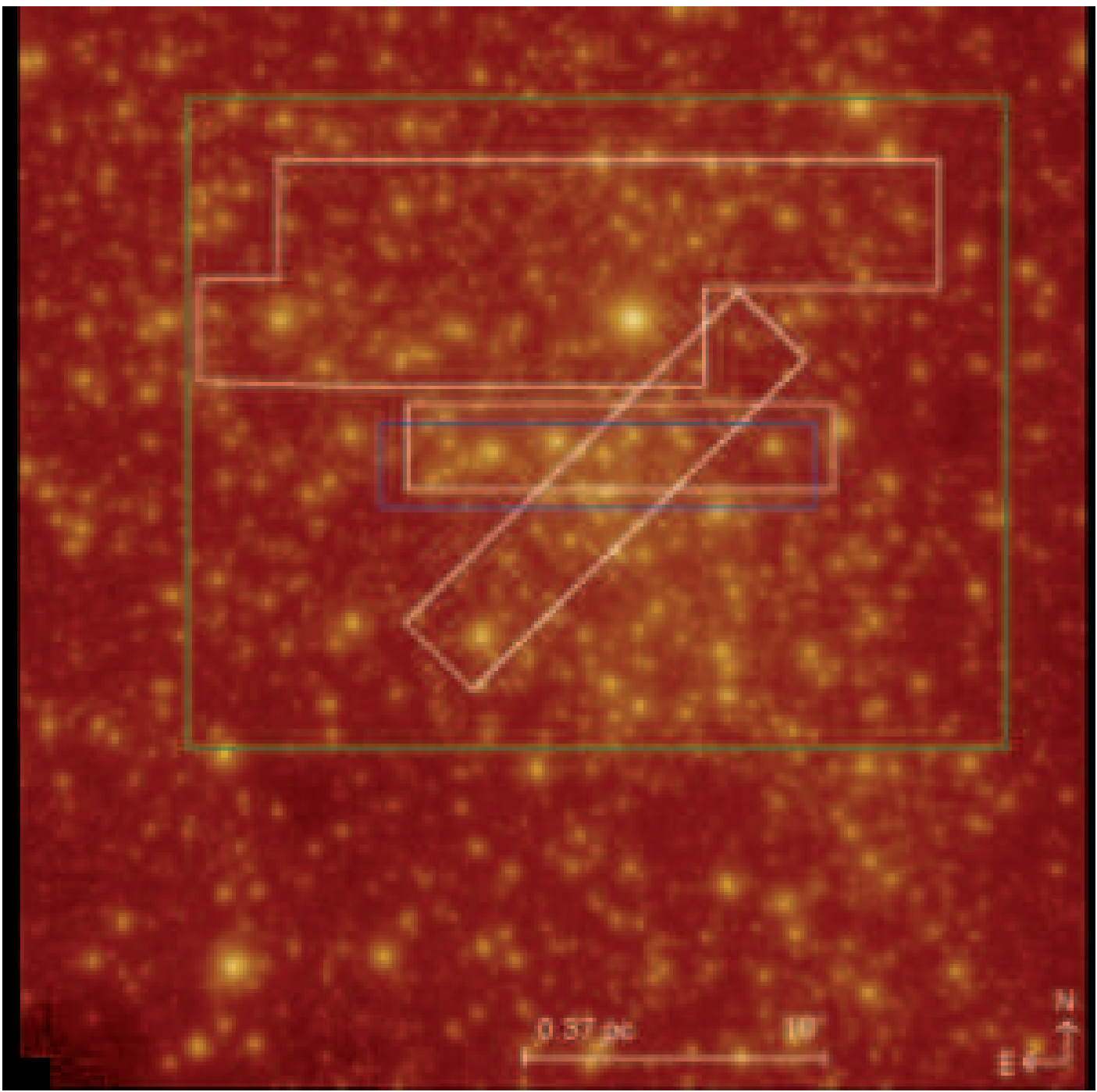}
\caption{\small Ks-band image of the innermost 40.5'' of the GC \citep[ESO VLT NACO image, 2004, see][]{buchholz2009}.
The marked areas denote the polarimetric FOVs: white indicates Ks-band, green Lp-band and blue H-band. The H-band (H1, see Tab.\ref{TabObservations} and the two southern
Ks-band FOVs (K1, K2) have been presented in \cite{buchholz2011}, while the northern Ks-band (K3) and the Lp-band (L1) FOV are examined in this study.}
\label{FigFOVwoll}
\end{figure*}
The data presented here were obtained using the NAOS-CONICA \citep[NACO, see][]{lenzen2003,rousset2003} instrument at the ESO VLT unit telescope 4
on Paranal in March 2011 (program 086.C-0049A, see Tab.\ref{TabObservations}).
The seeing varied during the observations, especially affecting the Ks-band dataset and the second night of Lp-band
observations. Only Lp-band observations taken in the first night were used. Using the infrared wavefront sensor
installed with NAOS, the bright super-giant IRS 7 located about 6'' north of Sgr A* was used to close the feedback loop
of the adaptive optics (AO) system. In the Ks-band, the sky background was determined by taking several dithered exposures
of a dark cloud 713'' west and 400'' north of Sgr A*, a region largely devoid of stars. The Lp-band sky was estimated
on a region located 60'' west and 60'' north of Sgr A*, switching between the sky and the target every other exposure.\\
We used the Wollaston prism available with NACO in combination with a rotatable half-wave plate for the polarization
measurements. The two channels produced by the Wollaston prism (0$^{\circ}$ and 90$^{\circ}$), combined with two orientations
of the half-wave plate (0$^{\circ}$ and 22.5$^{\circ}$), yielded four sub-images for each of several dither positions.
Each sub-image covered a field-of-view of 3.2''$\times$13.6'' (Ks-band), respectively 3.2''$\times$27.8'' (Lp-band), due to the different
optics that were used (see Tab.\ref{TabObservations}). 26 of these sub-images were used in the Lp-band, while the low data quality
in the Ks-band forced us to discard 26 sub-images in the 0$^{\circ}$/90$^{\circ}$ channel (leaving 21 out of 47 taken in total), as well
as 23 out of 47 in the 45$^{\circ}$/135$^{\circ}$ channel (leaving 25). In total, we were able to cover a field-of-view of 8''$\times$25'' (Ks-band),
respectively 26''$\times$28'' (Lp-band).
This corresponds to 0.30 pc $\times$ 0.96 pc and 0.96 pc $\times$ 1.04 pc, respectively (see Fig.\ref{FigFOVwoll}). Per sub-image, the area
covered corresponds to 0.12 pc $\times$ 0.51 pc (Ks-band) and 0.12 pc $\times$ 1.04 pc.\\
All images were corrected for dead/hot pixels, sky subtracted and flat-fielded. It is essential for a correct calibration that
the flat-field observations are taken through the Wollaston, because the flat-field shows variations caused
by different transmissivity of the channels as well as effects of the inclined mirrors behind the prism
\citep[see][for details on these effects]{witzel2010,buchholz2011}.\\
\begin{figure}[!t]
\centering
\includegraphics[width=\textwidth,angle=-90, scale=0.32]{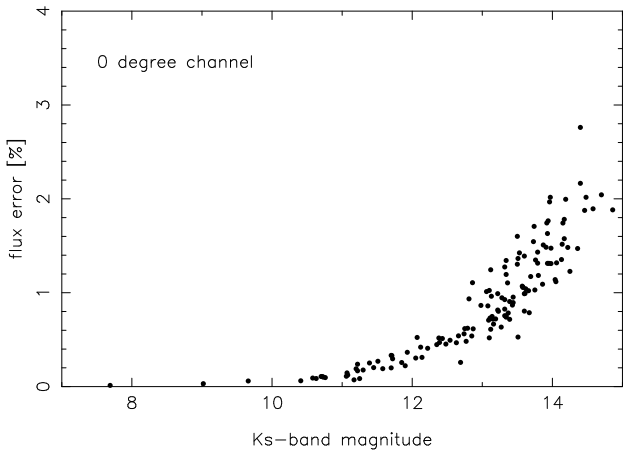}
\caption{\small Flux uncertainties for the 0 degree channel of the Ks-band dataset, plotted against the brightness of the
sources in magnitudes. The
uncertainties were determined from the flux variations in background apertures and can therefore only be regarded as a lower
limit for the actual uncertainties. The other channels show a very similar uncertainty distribution.}
\label{FigFluxerrKs}
\end{figure}
Even after the sky-subtraction and flat-fielding, the March 2011 Lp-band images still contained significant
patterns. These are not actual structures in the GC itself
\citep[as a comparison with previous Lp-band observations reveals, see e.g.][]{viehmann2005}, but must have been introduced either by the detector itself or possibly by the sky correction. This can happen when the sky
exposures contain sources themselves. Due to the high sky flux, these cannot be easily made out and masked, as
it can be done in the Ks-band. Two distinct patterns occur in all images: a series of 'stripes' along the
East-West-axis and  a 'cloudy' structure in the east and west of the FOV. The latter occurs at the same position
in each image, while the former is different for each exposure.\\
The East-West pattern was approximated by averaging over the x-axis of each image (excluding bright pixels, in order
to avoid a bias from the stars). The resulting profile was subtracted from the image. This does not introduce a significant bias, since there are no large scale East-West structures expected in this FOV.\\
In order to remove the stationary pattern, a median image was computed from all individual exposures (excluding
the stellar sources) for each polarization channel. This yielded a characteristic pattern for each channel, which
was then subtracted from all images.\\
Before and after the removal of each pattern, each image was shifted to a background level centered around zero.
This turned out to be necessary since the background level after sky subtraction varied on the order of 10-20
counts between
individual images. With a highly variable sky (due to less-than-optimal observing conditions), this can be expected:
the flux from the sky alone reaches 2000-3000 counts per pixel, so a change of 1\% between sky and object exposure
already produces an offset of 20-30 counts. Unfortunately, this means that the background flux cannot be measured
and any information about the background polarization is lost.\\ 
Compared to the data presented in \cite{buchholz2011}, our new observations cover a much greater FOV, at the price of a lower
depth. The low data quality of the Ks-band data further limits this study to the brighter sources in the field. 
\subsection{Photometry}
\subsubsection{Ks-band}
Unlike the data contained in our previous study \citep{buchholz2011}, where all individual images covered essentially the same FOV
in the very center of the GC, the data taken in March 2011 sacrificed depth in order to achieve a much greater FOV. This made
photometry on a combined image impractical, because the overlapping regions of the images were small, which in turn led to a
variable depth over the field.\\
In addition, the observations suffered from a small isoplanatic angle. Therefore, the PSF changed visibly across the FOV.
This effect varied in strength between the individual exposures. About half
of the Ks-band exposures had to be discarded because the very low data quality did not allow reliable photometry. This introduced
a further complication: about 50\% of the individual images had to be discarded because of insufficient quality, and this left very
few corresponding pairs of 0$^{\circ}$/45$^{\circ}$ images (covering the same FOV and taken in sequence). The different data
quality can also introduce offsets
in total intensity between the exposures (due to variable AO performance), and this made it necessary to conduct the photometry on each
exposure individually.\\
For this dataset (as well as for the Lp-band data, see below), using PSF-fitting or deconvolution-assisted photometry 
\citep{buchholz2009,schoedel2010a,buchholz2011} proved impractical.
The guide star IRS~7 was only covered in four of the selected exposures, and while this star would provide an excellent estimate of the
PSF including the faint wings, the variation of the shape of the PSF between the different images did not allow only this PSF to be used
for all other images as well. Local PSF extraction is complicated by the small number of suitable PSF stars in each exposure. The large
residua that are produced by using a poorly fitting PSF in this case outweigh any advantage of these otherwise much more precise
photometric methods.\\
\begin{figure}[!t]
\centering
\includegraphics[width=\textwidth,angle=-90, scale=0.32]{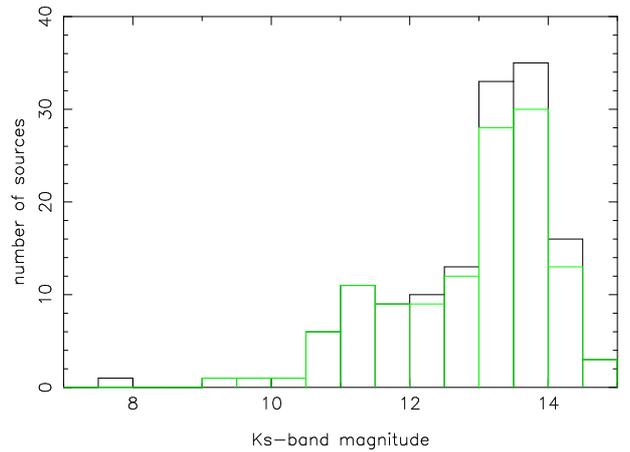}
\caption{\small Ks-band magnitudes of the sources detected in the polarimetric dataset K3. Black: all sources. Green: sources with
reliably measured polarization parameters.}
\label{FigKbandmags}
\end{figure}
We therefore determined that the best way to achieve reliable photometry on the individual images was aperture photometry, using
manually placed apertures for all sources in the FOV. Due to the low depth of the observations, the issue of crowding is far less
important than in our previous study \citep{buchholz2011}, especially since precise photometry is not possible here for fainter
sources (15-16 mag) anyway.\\    
\begin{figure}[!t]
\centering
\includegraphics[width=\textwidth,angle=-90, scale=0.35]{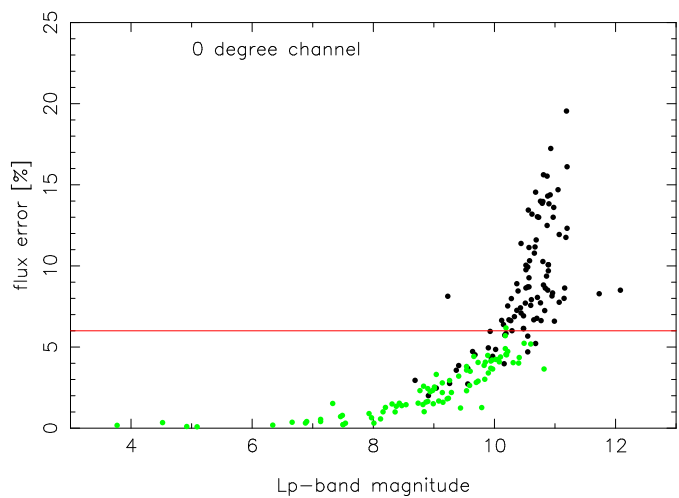}
\caption{\small Flux uncertainties for the 0 degree channel of the Lp-band dataset, plotted against the brightness of the sources
in magnitudes. The green points denote sources with reliable polarization parameters as was determined later in the analysis, while
the black points indicate sources that were rated as unreliable.
The uncertainties were determined from the flux variations in background apertures and can therefore only be regarded as lower limit for the
actual uncertainties. The other channels show a very similar uncertainty distribution.}
\label{FigFluxerrLp}
\end{figure}
The apertures were created based on a mosaic image, by placing an aperture on each discernible source and adjusting its radius so
that the whole PSF of each source was covered. Since only sources brighter than $\sim$14 mag were used (see below), source crowding was
not an issue, and contamination by the flux of neighboring or faint background sources can be assumed to be minimal. Brighter sources
were covered by larger apertures, on the order of 1'' for the two brightest sources in the field, IRS 7 (and IRS 3 in Lp-band).
The apertures for the fainter sources were significantly smaller, on the order of 0.2''.\\
Using these apertures, aperture photometry was conducted on the sources present in each image. Only apertures fully contained in the respective FOV of each image were used (in order to avoid problems
with sources not fully contained in the FOV). The background for each source was estimated using an average over four background
apertures close to each source, which were placed in regions with no visible sources. This background flux was then subtracted from
the flux measured for the source. \\
The resulting fluxes were used to calculate the Stokes Q and U parameters for each pair of 0$^{\circ}$/90$^{\circ}$ and 
45$^{\circ}$/135$^{\circ}$ exposures (see \S \ref{SectPolarimetry}). These values were then averaged to one Q and U value for each source.
This eliminates the bulk of the offset in flux between individual exposures. No systematic offsets in the Q or U parameter were found
between exposures.\\
The uncertainties of the measured fluxes were determined by calculating the FWHM of the distribution of the fluxes in the
background apertures. The value
\begin{equation}
\label{EqSigmaFWHM}
\sigma = \frac{FWHM}{2 \sqrt{2 ln(2)}}
\end{equation}
was adopted as the flux uncertainty per pixel. This led to a total flux uncertainty for a given aperture of
\begin{equation}
\label{EqSigmaAper}
\sigma_{aper} = \sqrt{2 n} \times \sigma
\end{equation}
, with $n$ as the number of pixels contained in the aperture. These flux uncertainties contributed one component to the uncertainties calculated for
Q and U (see below). Fig.\ref{FigFluxerrKs} shows these flux uncertainties in relation to the measured fluxes. The uncertainty distribution shows
a clear dependency on the flux, which can be expected since this is basically a Poisson uncertainty. This only sets a lower limit for the actual
uncertainty of the measured Q and U parameters. The main contribution to the uncertainties that were determined for these values stems from the 
variation over several exposures (see below).\\ 
\begin{figure}[!t]
\centering
\includegraphics[width=\textwidth,angle=-90, scale=0.32]{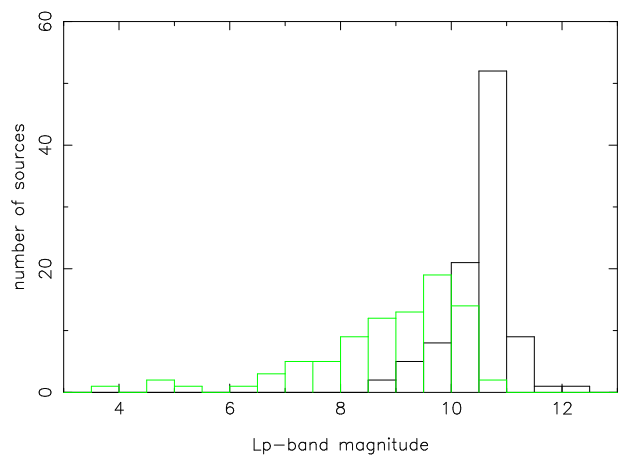}
\caption{\small Lp-band magnitudes of the sources detected in the polarimetric dataset L1. Black: all sources. Green: sources with
reliably measured polarization parameters.}
\label{FigLbandmags}
\end{figure}
This technique was used to measure the fluxes of the 143 visible sources in the observed FOV of 8.6''$\times$25''. Compared to the Ks-band data presented
in \cite{buchholz2011}, the number of detected sources per arcsec$^2$ is lower by a factor of $\sim$6. This illustrates the much lower
depth and worse data quality of the 2011 data.\\
In order to determine the depth of the new Ks-band observations, we used the Ks-band magnitudes provided for the observed
sources in \cite{schoedel2010b}. Fig.\ref{FigKbandmags} shows the resulting Ks-band luminosity function. Accoridngly, we estimate the
depth of the new observations as $\sim$14 mag, which is shallower by 1.5 mag than the data used in \cite{buchholz2011}.\\  
\subsubsection{Lp-band}
The Lp-band images varied much less in quality than the Ks-band dataset, and all exposures could be used for photometry. The small
overlap between the individual images and the fact that most of the FOV was only covered by a single exposure made it necessary to
conduct the photometry on the individual images. This was done in a similar way to the Ks-band dataset (see above): the sources in
the FOV were covered by manually placed apertures, while the background was determined in a 2 pixel annulus around each aperture.
Since the background was shifted to an average of zero for all images before (see \S \ref{SectObservations}), the residual background
only contributed a small amount to the total flux measured in each aperture.\\
Unlike in the Ks-band dataset, each FOV was covered by two corresponding 0$^{\circ}$ and 45$^{\circ}$ images. This allowed a direct 
comparison of the total intensity measured for each source by adding up the fluxes in the 0$^{\circ}$ and 90$^{\circ}$ respectively the
45$^{\circ}$ and 135$^{\circ}$ channels. For each image pair, the total intensities of all sources were added and a relative calibration 
factor was determined:
\begin{equation}
c_{totalint} = \frac{\sum_i (f_{00,i} + f_{90,i})}{\sum_i (f_{45,i} + f_{135,i})}.
\end{equation}
All fluxes measured in the 45$^{\circ}$ and 135$^{\circ}$ channels were multiplied with this factor in order to ensure the same relative 
flux calibration in all corresponding images. Due to the small overlaps between the images (the two northernmost FOVs showed no overlap
at all), a cross-calibration over the full FOV was not possible in this way. Therefore, the measured fluxes were normalized to an average
of one for each source before combining them to a common list of all detected sources. This ensures that possible differences in flux
for sources detected in more than one exposure did not influence the polarization determined for that source.\\
In order to determine the depth of the observations, we used the Lp-band magnitudes given for most of the observed sources in
\cite{schoedel2010b}. The few remaining sources that were not contained in this catalogue were calibrated by comparing their
measured fluxes with those of the sources with existing reference values in \cite{schoedel2010b}.\\  
The flux uncertainties per pixel were determined from the FWHM of the distribution of all background pixels of each image (see
Eq.\ref{EqSigmaFWHM}), and this value was used to calculate a total flux uncertainty for each source (see Eq.\ref{EqSigmaAper}).
Fig.\ref{FigFluxerrLp} shows these uncertainties in relation to the brightness of the sources, and they show the expected distribution of a
Poisson uncertainty. Only sources with a relative flux uncertainty of less than 6\% were used in the subsequent analysis. This effectively sets a
lower brightness limit of $\sim$10.5 mag (see Fig.\ref{FigLbandmags}) for a reliable determination of the polarization parameters. 
\subsection{Polarimetry}
\label{SectPolarimetry}
We determined the polarization degree and angle of each source by converting the measured normalized fluxes
into normalized Stokes parameters:
\begin{eqnarray}
I &=& 1\nonumber\\
Q &=& \frac{f_{0}-f_{90}}{f_{0}+f_{90}}\\
U &=& \frac{f_{45}-f_{135}}{f_{45}+f_{135}}\\
V &=& 0\nonumber
\end{eqnarray}
Since NACO is not equipped with a $\frac{\lambda}{4}$ plate, it was not possible to measure circular polarization.
But since the circular polarization of stellar sources in the GC is at best very small \citep{bailey1984}, we
assume here that it can be neglected and set to 0 at our level of accuracy. Note that this may not be the case
for dusty sources where circular polarization may result from multiple scattering if the local dust density is high enough.
Polarization degree and angle can then be determined in the following way:
\begin{eqnarray}
P &=& \sqrt{Q^2+U^2}\\
\theta &=& 0.5 \times atan \left( \frac{U}{Q} \right)
\end{eqnarray}
Uncertainties for Q and U (and subsequently p and $\theta$) were determined by error propagation of the measured flux uncertainties.
Since this just yields a lower limit for the true uncertainty, the standard deviation of the Q and U parameters of the sources that were
found in more than one field was added quadratically to the uncertainties determined from the flux uncertainties alone.\\    
In order to check whether or not the polarization of a source was determined reliably, we used a combination of
several methods: in the Lp-band, we calculated the normalized fluxes that would be expected for the determined values of
p and $\theta$. The difference between these values and the measured fluxes was then compared to the photometric uncertainties
of each data-point. The source was only classified as reliable if the root-mean-square of the deviations did not exceed
the root-mean-square of the relative photometric uncertainties. In addition, sources with $dp > p$ (the uncertainty of the polarization
exceeding its value) were rated as unreliable. Due to the selection issues in the Ks-band, the first method was not applicable.
Here, only the latter method was used, with sources with $dp > p$ rated as unreliable measurements. The subsequent analysis
is only based on the sources classified as reliable.\\
Using the analytical model for the polarimetric calibration of NACO developed by \cite{witzel2010}, we were able to
reduce the systematic uncertainties produced by instrumental polarization and achieve a direct calibration for this
larger FOV as well \cite[compared to the FOV presented in][thereby expanding our previous results]{buchholz2011}. This method
reduces the systematic uncertainties
of polarization degrees and angles to $\sim$1\% and $\sim$5$^{\circ}$, so our final values are dominated by photometric
uncertainties.
\subsection{Correcting for foreground polarization}
\label{SectForegroundremoval}
In the same way as in \cite{buchholz2011}, we assume here that the total effect of the foreground polarization can be treated
as a linear polarizer with a certain orientation $\theta_{fg}$ and efficiency $p_{fg}$. This can be described by a Mueller matrix,
\begin{equation}
S_{obs} = M_{rot}( - \theta') \times M_{lin}(p) \times M_{rot}(\theta') \times S_{int},
\end{equation}
with $S_{obs}$ as the observed total Stokes vector and $S_{int}$ as the Stokes vector of the intrinsic polarization.
The Mueller Matrix $M_{lin}(p)$ describes a linear polarizer, producing a maximum of polarization along the North-South-axis.
This matrix has to be rotated to the appropriate angle by multiplying it with $M_{rot}(\theta')$, a standard 4$\times$4 rotation
matrix. The angle $\theta' = 90^{\circ} + \theta_{fg}$ has to be used in the rotation matrix, since we define the polarization angle $\theta$ as
the angle where we measure the flux maximum, while the angle of reference for the Mueller
matrix describing the linear polarizer is the angle where the maximum in absorption occurs. The matrices are given in \cite{buchholz2011}.\\
For each source to which the depolarization matrix was applied, the foreground polarization had to be estimated. The accuracy of this
estimate controls the relevance of the resulting intrinsic polarization. We used the average of the polarization parameters of the
surrounding point sources as an estimate for $\theta_{fg}$ and $p_{fg}$.\\ 
The resulting matrix can then be inverted and multiplied with the calibrated observed Stokes vector of a source to
remove the foreground polarization and leave only the intrinsic polarization. We applied this method to the extended sources, both
to their total polarization and the polarization maps, in order to isolate their intrinsic polarization pattern (see \S \ref{SectExtended}).
For each extended source, we indicated the point sources used as foreground references(see subsections in \S \ref{SectExtended}). 
\section{Results and discussion}
\label{SectResults}
\subsection{Ks-band polarization}
\label{SectKpol}
\begin{figure*}[!t]
\centering
\includegraphics[width=\textwidth,angle=0, scale=1.0]{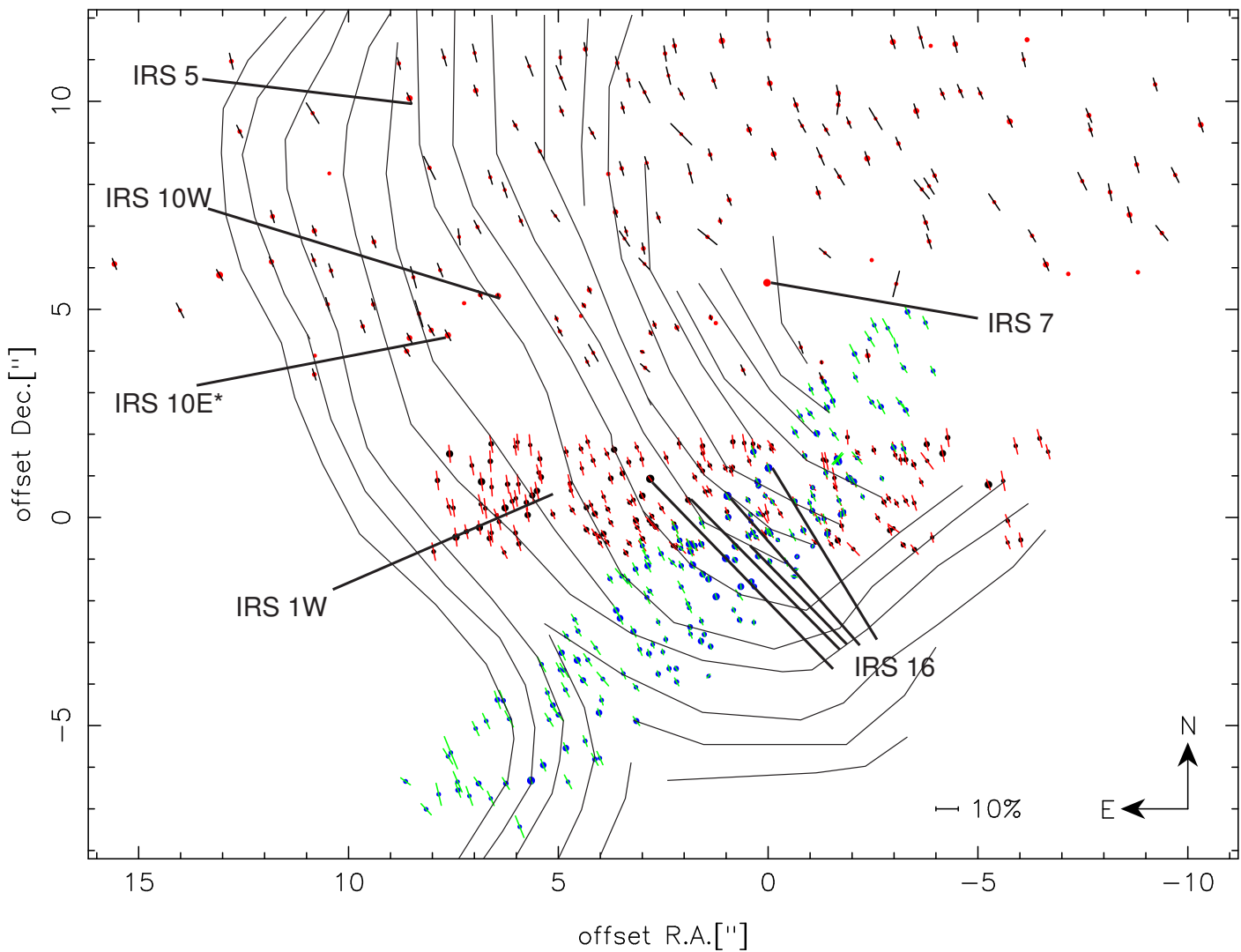}
\caption{\small Ks-band polarization map of stars in the Galactic center. Only reliably measured polarization values are shown
here. Black circles / red lines and blue circles / green lines: datasets presented in \cite{buchholz2011}. Red circles and black
lines: new March 2011 Ks-band data. The diameter of the circles corresponds to the brightness of the source.
Thin black lines in the background denote magnetic fields determined from MIR data \citep[][based on a 1.5'' beam]{aitken1998}.
The brightest sources are also indicated.}
\label{FigPolmapK}
\end{figure*}
\begin{figure*}[!t]
\centering
\includegraphics[width=\textwidth,angle=-90, scale=0.72]{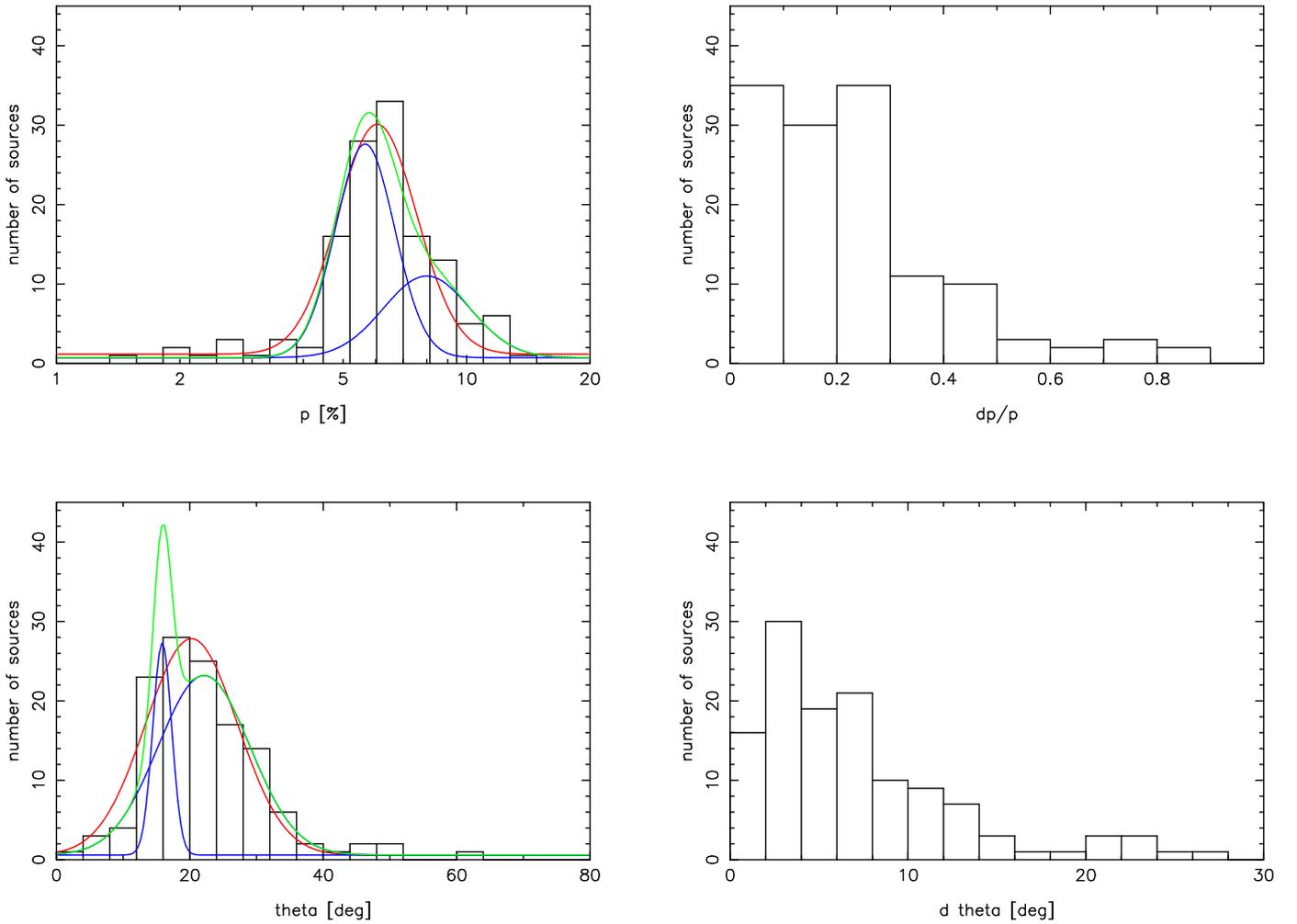}
\caption{\small Ks-band polarization degrees (plotted on logarithmic scale, upper left) and angles (lower left) of stars in
the Galactic center. The red lines denotes fits with a single Gaussian distribution, while the green
respectively blue lines denote the fit with a double Gaussian (green: sum, blue: individual Gaussians). Upper right: relative uncertainties
of the polarization degrees. Lower right: absolute uncertainties of the polarization angles.}
\label{FigPoldegsKnorth}
\end{figure*}

Despite the much wider FOV compared to \cite{buchholz2011}, the polarization parameters of only 126 sources could be measured reliably.
This still represents an increase by a factor of 4 compared to the sources in the same FOV for which \cite{eckart1995} determined the polarization
($\sim$ 30 sources in this area, while \cite{ott1999} did not cover this region at all).\\
As in the deeper datasets, the polarization angles show the expected alignment with the Galactic plane (see Fig.\ref{FigPolmapK}).
There may be a trend to larger polarization degrees towards the north compared to the southern center, but it is not as clear as the trend
found in the \cite{buchholz2011} data.\\
Fitting the distribution of the polarization angles with a single Gaussian yields a peak at 20$^{\circ} \pm$ 7$^{\circ}$ (see
Fig.\ref{FigPoldegsKnorth}, lower left frame). For comparison, the distribution was also fitted with a double Gaussian, which led
to two peaks at 16$^{\circ} \pm$ 1$^{\circ}$ and 22$^{\circ} \pm$ 7$^{\circ}$. This improves the reduced $\chi^2$ by a factor of 3,
but the fact that the secondary peak lies within the FWHM of the primary peak reduces the confidence in this feature.\\
The polarization degrees seem to show lower values towards the southern-central part of the FOV, which would link up well with the
results found in \citep[][see Fig.\ref{FigPolmapK}]{buchholz2011}. Fitting the logarithms of the polarization degrees with
a single Gaussian yields a peak at (6.1 $\pm$ 1.3)\%, while using a double Gaussian resulted in a lower reduced $\chi^2$ (by a
factor of 2), with peaks fitted at (5.7 $\pm$ 1.0)\% and (8.0 $\pm$2.1)\%.\\ 
The uncertainties of the polarization angle mostly stay below 15$^{\circ}$ (see Fig.\ref{FigPoldegsKnorth}, lower right frame), and this
further limits the confidence in the second peak due to its small offset from the primary feature. The relative uncertainties of the
polarization degree reach up to 50\% (except a few outliers, see Fig.\ref{FigPoldegsKnorth}, upper right frame).\\
A complete list of all detected sources with reliably measured polarization parameters is included as {\em online} material (table A1).
This list contains the position offsets from Sgr A* (RA, Dec), H-, Ks-, and Lp-band (see \S \ref{SectLpol}) polarization parameters of the sources observed in this work and in our previous
publication, \cite{buchholz2011}.
\subsection{Lp-band polarization}
\label{SectLpol}
\begin{figure*}[!t]
\centering
\includegraphics[width=\textwidth,angle=-90, scale=0.74]{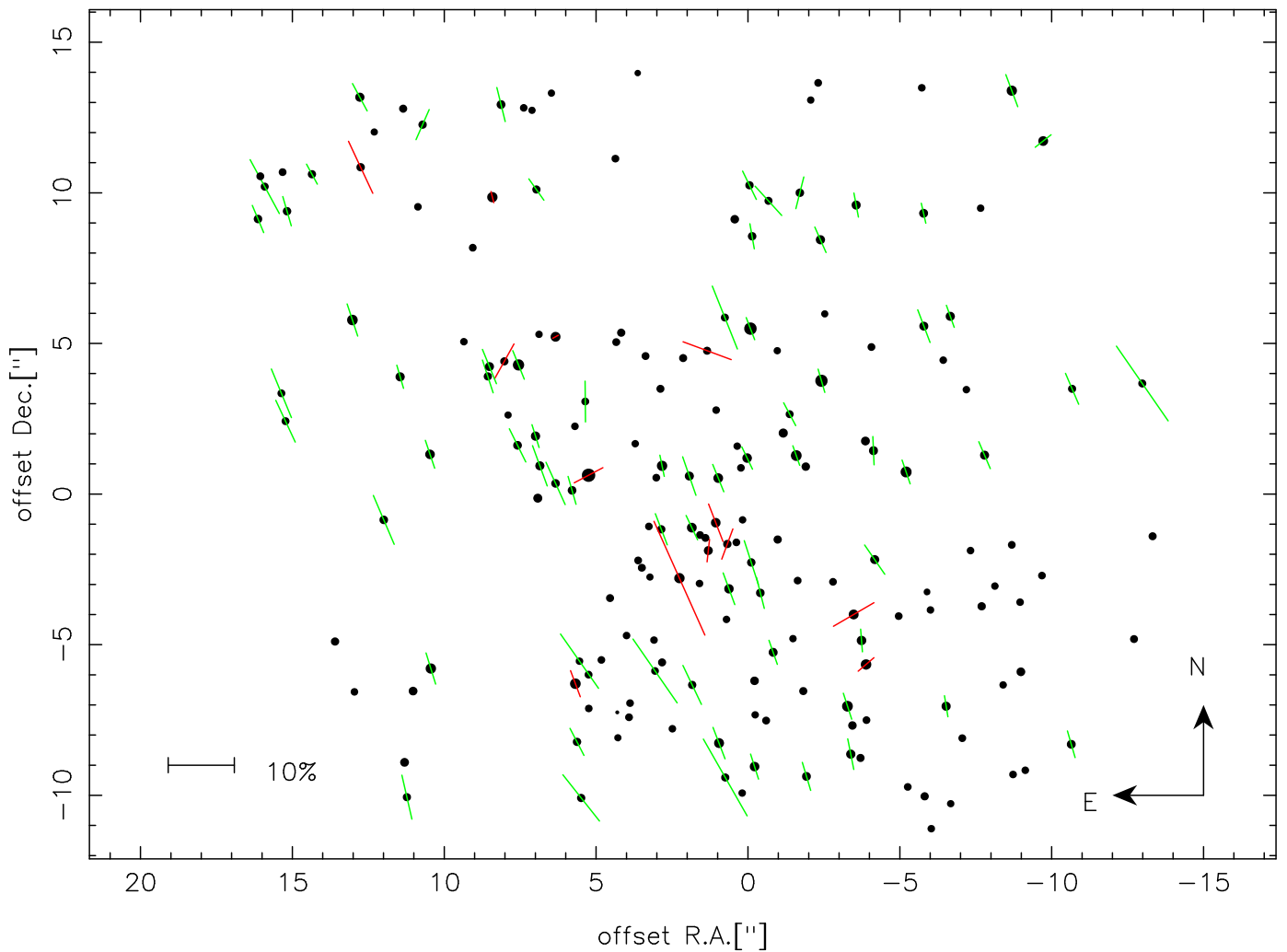}
\caption{\small Lp-band polarization map of stars in the Galactic center. Only reliably measured polarization values
are shown here. Red lines indicate the polarization vectors of sources classified as affected by intrinsic polarization according to
\S \ref{SectKLrelation} as well as the observed bow-shock sources, while green lines
denote sources only exhibiting foreground polarization.}
\label{FigPolmapL}
\end{figure*}
\begin{figure*}[!t]
\centering
\includegraphics[width=\textwidth,angle=-90, scale=0.72]{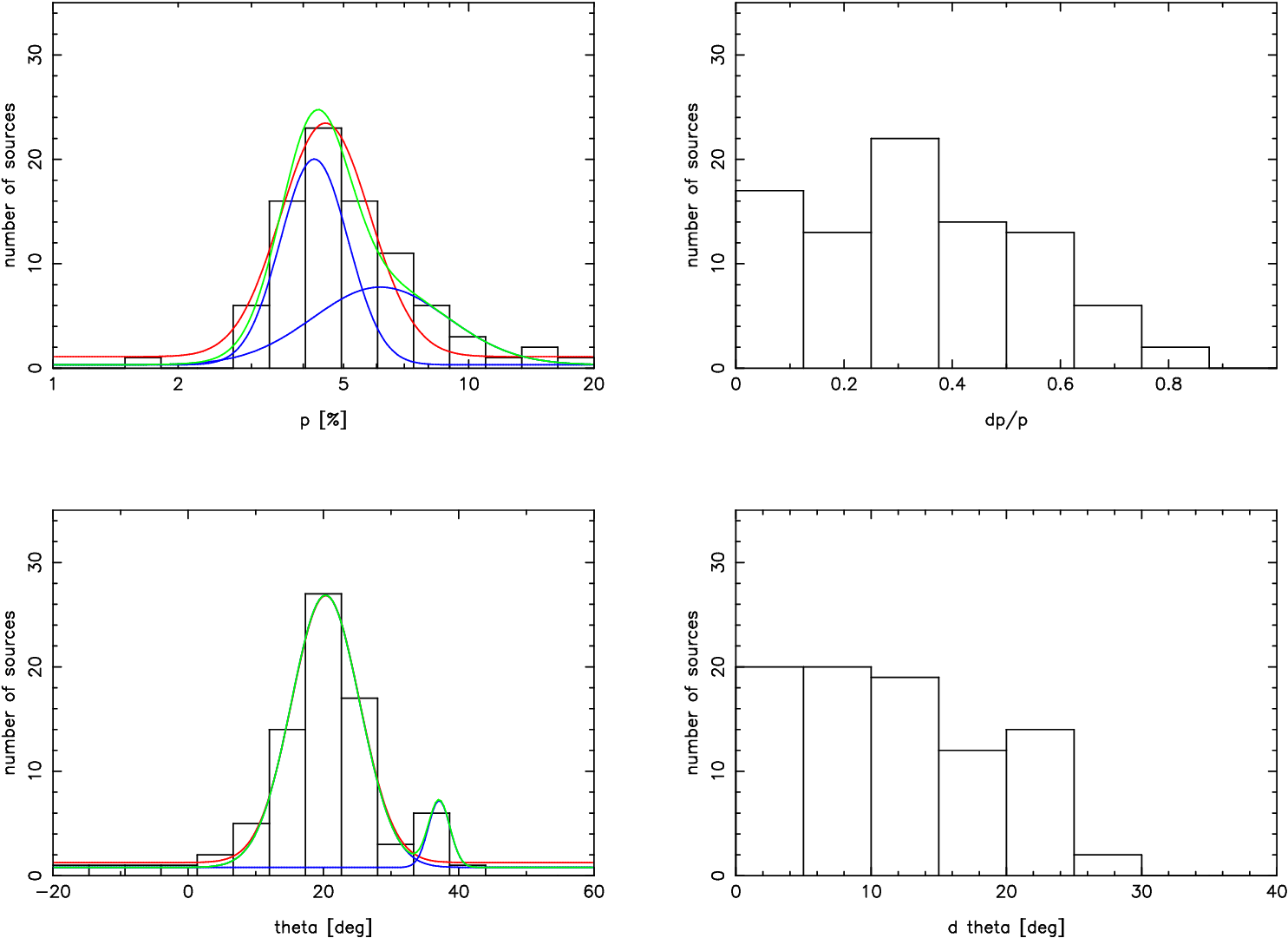}
\caption{\small Lp-band polarization degrees (plotted on logarithmic scale, upper left) and angles (lower left)
of stars in the Galactic center. The red line denotes the fit with one Gaussian distribution, while green respectively blue lines denote the
fit with a double Gaussian (green: sum, blue: individual Gaussians). Upper right: relative uncertainties of the polarization degrees. Lower right: absolute
uncertainties of the polarization angles.}
\label{FigPoldegsL}
\end{figure*}

The polarization of 84 sources in the Lp-band dataset could be determined reliably. Considering the size of the FOV, this number appears
small compared to the H- and Ks-band data presented in \cite{buchholz2011}, but Lp-band polarimetry suffers from much larger difficulties
than what is encountered at
shorter wavelengths. Fig.\ref{FigPolmapL} shows that while there is an overall trend of polarization along the Galactic plane (as seen
in the H- and Ks-band), there is a significant number of outliers with either strongly deviating polarization angles or much
weaker/stronger polarization degrees than what is expected for pure foreground polarization. Considering the complex structures observed
in the Lp-band in the central parsec (large-scale dust structures, bow-shocks, embedded sources), this is not surprising.\\
Despite these additional complications, the distribution of polarization angles (see Fig.\ref{FigPoldegsL}, lower left frame) shows a
clear peak at 20$^{\circ} \pm $5$^{\circ}$ when fitted with a single Gaussian. A small secondary feature might be present at
$\sim$35$^{\circ}$, and a fit with a double Gaussian indeed has a better reduced $\chi^2$ (by a factor of 2.5). But considering the
small number of sources contained in this feature compared to the number of the other, more widely distributed outliers, this may not
be a significant feature at all. In addition, the single Gaussian fits the distribution with sufficient accuracy, and it does not differ
much from the first peak of the fitted double Gaussian.\\
In the same way, the distribution of the logarithms of the polarization degrees shows a slightly widened peak at (4.5 $\pm$ 1.4)\% (see
Fig.\ref{FigPoldegsL}, upper left frame), which can also be fitted with a double Gaussian for a better reduced $\chi^2$ (by a factor of
5). This produces a secondary peak at $\sim$6\%, but again, this may not be significant, considering the width of this peak and the fact that
it is within the FWHM of the fitted single peak. The single Gaussian provides a sufficient fit.\\
Fig.\ref{FigPoldegsL}, upper right frame, shows the relative uncertainties of the polarization degree. These uncertainties mostly stay below $\sim$60\%,
a value significantly larger than what was found for the relative uncertainties of the other datasets \citep[see][]{buchholz2011}. This again indicates the problems
of Lp-band polarimetry (difficult photometry combined with lower polarization degrees compared to the H/Ks-band).
The uncertainties of the polarization angles (see Fig.\ref{FigPoldegsL}, lower right frame) reach up to $\sim$25$^{\circ}$, which is
also higher than what was found in the other datasets. The uncertainties of the polarization parameters further decrease the confidence in
the significance of the small secondary features found in both distributions.
\subsection{Comparison to previous results}
\label{SectComparisonPrev}
Our own recent study \citep{buchholz2011} was the first that measured the NIR polarization of GC sources at this spatial resolution. The FOV of that
work does not overlap with the new data, so a direct source-by-source comparison is not possible. The distributions of the polarization
parameters show similar features, however: the distribution of the angles found here (see Fig.\ref{FigPoldegsKnorth}) does not differ
much from what was found for datasets K1 and K2, and the fitted single Gaussian peaks agree within the uncertainties. The double
Gaussian fit is also consistent with dataset K1: the two peaks fitted here match the values found for the
peaks in dataset K1 within one respectively two sigma (peaks at 10$^{\circ}$/16$^{\circ}$ respectively 26$^{\circ}$/22$^{\circ}$).
On average, the polarization degrees are slightly higher than those found in our previous study (see Fig.\ref{FigPoldegsKnorth},
upper left frame), but the peak of the fitted single Gaussian still agrees with the value found in \cite{buchholz2011} within the
uncertainties. The peaks of the fitted double Gaussian are also consistent with the two peaks fitted in our last study, but the second
feature found here is a lot less pronounced compared to the previously presented data.\\
A comparison to the values presented in \cite{eckart1995} reveals considerable deviations: only 47\% of the 32 sources found both in the
older study and the new dataset show an agreement in polarization degree within 3$\sigma$ or less. 63\% of the sources show a 3$\sigma$
agreement for the polarization angles. The FOV of this dataset and that of \cite{ott1999} do not overlap, so no comparison was possible.
It has to be noted, however, that the agreement with these older studies was significantly better for the sources contained in
\cite{buchholz2011}.\\
In general, differences between these older studies and the new measurements can probably be attributed to the lower spatial resolution
of the former. But the lower data quality of the observations presented in this work compared to the data used in our previous study
most likely also reduces the agreement with the \cite{eckart1995} values.
Both older studies show average polarization angles generally parallel to the Galactic plane, on average at 25$^{\circ}$ respectively 
30$^{\circ}$. Similar results are found here for the much larger new sample of sources. The distribution of the polarization degrees
determined from our new data peaks at higher values than the flux-weighted average polarization degree of the older studies, but that
can be expected since the inclusion of IRS~7 with its polarization degree of only 3.6\% in both older surveys lowered the flux weighted
average that was calculated there considerably \citep[see also the note on flux-weighted averaging in][]{buchholz2011}.\\
There are only a few sources in the FOV of the presented dataset that can be compared directly to the K-band results of \cite{knacke1977}:
IRS~7 is bright and isolated enough to be detected as a single source even in the old data. IRS~3 would also be suitable for
such a comparison, but this source is not fully contained (i.e. located on the edge of the FOV) in any of the 2011 images.\\
\cite{knacke1977} measured a polarization of (3.0 $\pm$ 0.3)\% at 20$^{\circ} \pm$ 5$^{\circ}$ for IRS~7, and the value found here
matches that result quite well: (2.9 $\pm$ 1.0)\% at 13$^{\circ} \pm$ 10$^{\circ}$ have been measured based on the 2011 data, but
it has to be cautioned that the saturation of the source might influence these results.\\ 
In the Lp-band, there are even less possibilities for a comparison: \cite{lebofsky1982} (5.8'' beam) and \cite{knacke1977} (7'' beam)
are the only studies to date that measured L-band polarization in the GC. Only two sources can be compared directly: \cite{knacke1977}
measured the polarization of IRS~3 as (3.4 $\pm$ 1.0)\% at $16^{\circ} \pm 8^{\circ}$, while they give values of (2.6 $\pm$ 0.3)\% at
$43^{\circ} \pm 5^{\circ}$ for IRS~7. \cite{lebofsky1982} do not provide values for IRS~3, but measured (3.2 $\pm$ 0.5)\% polarization at
$13^{\circ}$ for IRS~7. The values found in our observations for these two sources, about 30 years and several instrument/telescope
generations later, agree relatively well: the polarization of IRS~3 is measured as (3.7 $\pm$ 0.1)\% at $16.0^{\circ} \pm 0.7^{\circ}$, 
which is in very good agreement with the \cite{knacke1977} value. IRS~7 exhibits a polarization of (3.6 $\pm$ 0.1)\% at $20.9^{\circ} \pm
0.7^{\circ}$. This matches the \cite{lebofsky1982} values better, while the measured angle deviates from \cite{knacke1977}. The
good agreement of these polarization parameters indicates that the recently developed calibration model of \cite{witzel2010}
is applicable to the Lp-band as well, though further observations of polarimetric standard sources in this wavelength band would be
desirable to support this.
\subsection{Detecting intrinsic polarization}
\label{SectIntrinsic}
The variation of the foreground polarization over the FOV \citep[see e.g.][and results presented above]{buchholz2011} makes it difficult
to determine exactly whether a source is affected by additional intrinsic polarization. If only one wavelength band is available, the only
possibility is a comparison of $Q$ and $U$ (or $p$ and $\theta$) for each source to the parameters found for the neighboring sources. In
case of deviations that exceed a given theshold, the source can be classified as intrinsically polarized. This threshold can be determined
from the width of the distributions of the parameters.\\
Another option exists where measurements in several wavelength bands are available as it is the case for the sources common to Ks- and Lp-band
(as shown in
\S \ref{SectKLrelation}). The relation between Ks- and Lp-band polarization of a source can provide a strong indication of intrinsic
polarization: if $\frac{p_{Lp}}{p_{Ks}}$ is significantly higher or lower than the value expected for foreground polarization
\citep[$\sim$0.7-0.8 towards the GC, see ][]{jones1990} or if the polarization angles deviate strongly between the two bands,
the source can be regarded as intrinsically polarized. The reason for this is that at least in the foreground component, the same
grains should be responsible for the polarization in all NIR (and optical) bands. While different grain parameters and alignments
could lead to different contributions at the observed wavelengths, such influences will generally even our along the LOS, and a purely
foreground-polarized source should show the same polarization angle over these wavelength bands. If it does not, another local component must
be contributing.\\
Tab.\ref{TabKLcomp} lists the sources detected in both bands, classifying them as mainly foreground affected or as candidates for
intrinsic polarization.
This was determined based on the comparison between the two wavelength bands (see \S \ref{SectKLrelation}). 
\begin{figure*}[!t]
\centering
\includegraphics[width=\textwidth,angle=-90, scale=0.335]{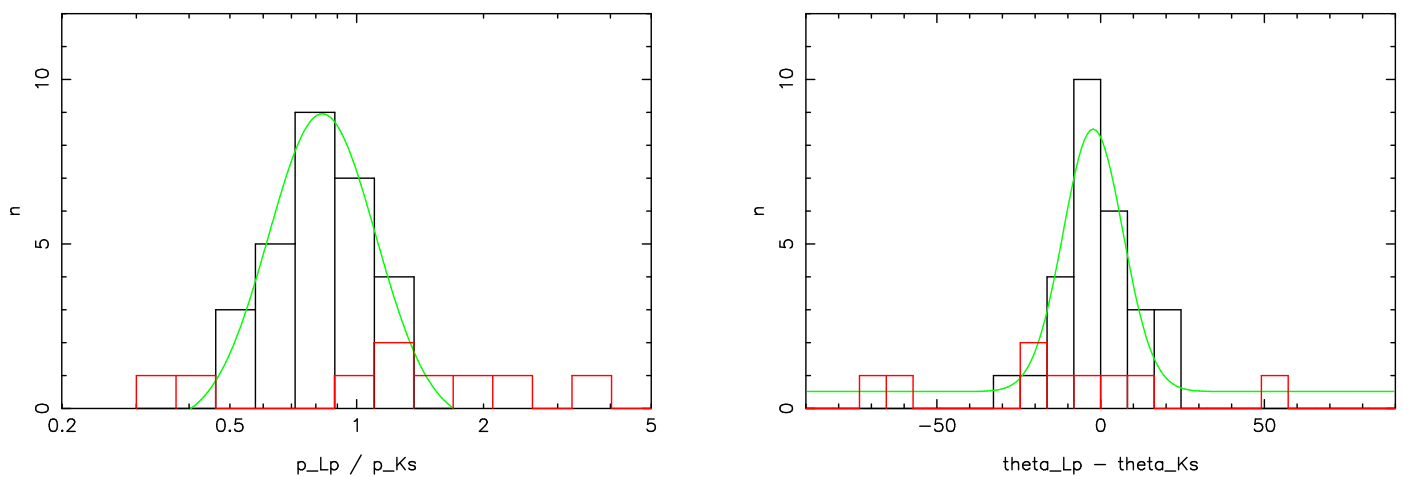}
\caption{\small Left frame: relation of Lp- to Ks-band polarization
degree (logarithmic plot). Right frame: difference between Lp- and Ks-band polarization angle. In both frames, the red columns
represent suspected intrinsically polarized sources, while the black columns contain sources affected only by the foreground
polarization. The green lines represent Gaussians fitted to the histograms.}
\label{FigPolcompKLtot}
\end{figure*}
\subsection{Relation between Ks- and Lp-band foreground polarization}
\label{SectKLrelation}
For 37 sources, Ks-band and Lp-band polarization parameters have been measured. This is less than half of the available Lp-band
sources, but that can be expected considering the respective fields-of-view observed in both bands (see Fig.\ref{FigFOVwoll}).
Of these common sources, 9 candidates for intrinsic polarization have been found based on the methods described in \S
\ref{SectIntrinsic}. Tab.\ref{TabKLcomp} shows the polarization parameters of the common sources, as well as the classification of
each source as intrinsically or foreground polarized. Several bright likely intrinsically polarized sources are examined in more detail in
\S \ref{SectExtended}. 
\begin{table*}[!t]
\caption[Polarization parameters for sources common to Ks/Lp-band]{\small Polarization parameters of the sources detected in the Ks-
and Lp-band. {\em ID} indicates the number of the source in the list contained in the {\em online} material. {\em Name} either lists the
GCIRS number of the source or the corresponding source number(RSxxxx) in \cite{schoedel2010b}. {\em Class} gives the classification of the source as intrinsically (int) or foreground (fg) polarized.} 
\label{TabKLcomp}
\centering       
\begin{tabular}{l l r r r r l}
\hline\hline
ID & name & $p_{Ks}$ [\%] & $p_{Lp}$ [\%] & $\theta_{Ks}$ [$^{\circ}$] & $\theta_{Lp}$ [$^{\circ}$] & class\\
\hline
1 & IRS~16NE & 2.7 $\pm$ 0.6 & 3.3 $\pm$ 0.6 & 35 $\pm$ 11 & 12 $\pm$ 8 & fg\\  
2 & IRS~16C & 4.3 $\pm$ 0.6 & 4.5 $\pm$ 0.9 & 25 $\pm$ 5 & 22 $\pm$ 7 & fg\\  
5 & IRS~1C & 8.2 $\pm$ 1.0 & 6.6 $\pm$ 1.6 & 8 $\pm$ 6 & 20 $\pm$ 9 & fg\\  
7 & IRS~1NEE & 7.8 $\pm$ 0.8 & 5.7 $\pm$ 2.3 & 5 $\pm$ 5 & 26 $\pm$ 14 & fg\\  
8 & IRS~1SE & 7.6 $\pm$ 0.8 & 7.1 $\pm$ 1.7 & 11 $\pm$ 5 & 24 $\pm$ 8 & fg\\  
11 & RS788 & 6.1 $\pm$ 0.8 & 4.4 $\pm$ 0.9 & 11 $\pm$ 5 & 16 $\pm$ 8 & fg\\  
12 & IRS~16NW & 3.9 $\pm$ 0.8 & 3.8 $\pm$ 1.1 & 26 $\pm$ 10 & 25 $\pm$ 10 & fg\\  
184 & IRS~34 & 7.5 $\pm$ 0.5 & 4.3 $\pm$ 2.4 & 11 $\pm$ 5 & 4 $\pm$ 24 & fg\\  
187 & IRS~6E & 4.8 $\pm$ 0.8 & 3.8 $\pm$ 0.5 & 19 $\pm$ 8 & 18 $\pm$ 5 & fg\\  
195 & IRS~9 & 2.9 $\pm$ 0.9 & 4.2 $\pm$ 0.5 & 26 $\pm$ 16 & 20 $\pm$ 5 & int\\ 
196 & IRS~16SW3 & 2.7 $\pm$ 0.9 & 3.5 $\pm$ 1.6 & 17 $\pm$ 17 & -6 $\pm$ 23 & int\\
197 & IRS~16SWE & 3.9 $\pm$ 1.0 & 4.1 $\pm$ 0.8 & 24 $\pm$ 9 & 25 $\pm$ 6 & fg\\  
199 & IRS~16SW2 & 2.1 $\pm$ 0.9 & 4.9 $\pm$ 3.0 & 4 $\pm$ 19 & -20 $\pm$ 21 & int\\
201 & IRS~35 & 6.2 $\pm$ 0.9 & 5.1 $\pm$ 3.1 & 26 $\pm$ 5 & 21 $\pm$ 22 & fg\\  
284 & IRS~16SW1 & 3.1 $\pm$ 0.7 & 6.1 $\pm$ 0.7 & 20 $\pm$ 14 & 21 $\pm$ 5 & int\\
300 & IRS~29 & 2.8 $\pm$ 0.8 & 3.2 $\pm$0.5 & 24 $\pm$ 11 & 19 $\pm$ 5 & fg\\  
305 & RS1378 & 5.0 $\pm$ 0.7 & 3.9 $\pm$ 2.8 & 25 $\pm$ 5 & 28 $\pm$ 25 & fg\\  
389 & RS3909 & 5.6 $\pm$ 0.5 & 5.2 $\pm$ 0.5 & 30 $\pm$ 5 & 18 $\pm$ 5 & fg\\ 
399 & IRS~5NE & 7.4 $\pm$ 0.5 & 8.8 $\pm$ 0.9 & 13 $\pm$ 5 & 25 $\pm$ 5 & int\\
400 & RS3078 & 5.2 $\pm$ 0.5 & 4.0 $\pm$ 3.1 & 17 $\pm$ 5 & 35 $\pm$ 23 & fg\\
401 & IRS~5 & 5.4 $\pm$ 0.5 & 1.8 $\pm$ 0.5 & 24 $\pm$ 5 & 14 $\pm$ 5 & int\\
403 & IRS~10W & 2.1 $\pm$ 0.5 & 0.9 $\pm$ 0.5 & 6 $\pm$ 5 & -61 $\pm$ 5 & int\\ 
404 & IRS~10E* & 5.4 $\pm$ 0.5 & 4.7 $\pm$ 0.5 & 28 $\pm$ 5 & 22 $\pm$ 5 & fg\\ 
405 & IRS~10E2 & 4.7 $\pm$ 0.5 & 5.6 $\pm$ 1.2 & 28 $\pm$ 5 & 22 $\pm$ 7 & fg\\ 
406 & IRS~10E1 & 5.3 $\pm$ 0.5 & 5.3 $\pm$ 2.7 & 32 $\pm$ 5 & 18 $\pm$ 17 & fg\\ 
407 & IRS~10E3 & 5.6 $\pm$ 0.5 & 6.1 $\pm$ 2.7 & 30 $\pm$ 5 & -29 $\pm$ 16 & int\\
411 & RS2257 & 6.4 $\pm$ 0.9 & 4.8 $\pm$ 2.4 & 15 $\pm$ 5 & 26 $\pm$ 17 & fg\\  
412 & RS2059 & 6.7 $\pm$ 0.5 & 6.0 $\pm$ 3.5 & 20 $\pm$ 5 & 42 $\pm$ 16 & fg\\  
413 & RS2214 & 7.5 $\pm$ 0.6 & 5.0 $\pm$ 3.1 & 12 $\pm$ 5 & -14 $\pm$ 24 & fg\\  
414 & RS2245 & 6.8 $\pm$ 0.5 & 3.8 $\pm$ 1.1 & 18 $\pm$ 5 & 11 $\pm$ 13 & fg\\  
421 & RS2538 & 5.5 $\pm$ 0.5 & 3.1 $\pm$ 1.3 & 22 $\pm$ 5 & 13 $\pm$ 19 & fg\\  
422 & RS1707 & 6.5 $\pm$ 0.5 & 4.3 $\pm$ 1.1 & 18 $\pm$ 5 & 24 $\pm$ 9 & fg\\  
424 & RS1644 & 5.5 $\pm$ 0.6 & 3.9 $\pm$ 2.0 & 19 $\pm$ 5 & 10 $\pm$ 23 & fg\\  
427 & IRS~7 & 2.9 $\pm$ 0.5 & 3.6 $\pm$ 0.5 & 13 $\pm$ 10 & 21 $\pm$ 5 & fg\\ 
432 & IRS~30W & 5.9 $\pm$ 4.9 & 3.6 $\pm$ 1.1 & 24 $\pm$ 10 & 19 $\pm$ 12 & fg\\  
515 & IRS~3 & 5.8 $\pm$ 4.8 & 3.7 $\pm$ 0.5 & 10 $\pm$ 15 & 16 $\pm$ 5 & fg\\  
526 & RS628 & 2.3 $\pm$ 0.5 & 7.8 $\pm$ 3.9 & 16 $\pm$ 8 & 70 $\pm$ 13 & int\\ 
\hline
\end{tabular}
\end{table*} 
Fig.\ref{FigPolcompKLtot} shows the distribution of $\frac{p_{Lp}}{p_{Ks}}$ and the difference between the polarization angles.
Most of the outlier values can be assumed to be affected by intrinsic polarization (plotted in red), while the distribution
of the foreground-polarized sources shows clear peaks at 0.8 $\pm$ 0.3 ($\frac{p_{Lp}}{p_{Ks}}$) respectively -2$^{\circ} \pm$
9$^{\circ}$ ($\theta_{Lp}-\theta_{ks}$). The latter value agrees with zero within the uncertainties, which is expected
if indeed the same grain population (or at least grains aligned in the same direction) is responsible for the polarization
at both wavelengths. The former exceeds the value expected from theoretical and semi-empirical models: while the \cite{serkowski1975}
law does not work well in the NIR anyway \citep[see e.g.][]{martin1990}, the model proposed by \cite{mathis1986} and the power law
relation suggested by \cite{martin1990} predict too small values as well. \cite{jones1990} compared these relations to data
obtained on local sources and observations on three GC sources, IRS~7, IRS~9 and GCS~9. The GC sources showed an even higher excess
than the local sources compared to the results of the models/relations, with $\frac{p_{Lp}}{p_{Ks}}$ around 0.7. This is confirmed
here for a much larger number of sources, while the cause remains uncertain. \cite{jones1990} suggested polarimetric observations
at even longer wavelengths and spectropolarimetric observations in the Lp-band as a possibility to shed some more light on this issue.\\
\cite{nagata1994} conducted such spectropolarimetric observations on several sources within 0.5$^{\circ}$ of the GC, including IRS 7,
finding a polarization excess over the whole 3$\mu$m window. They proposed that this excess might be a localized feature, and its source
not ubiquitous in the diffuse interstellar medium on galactic scales. It might prove interesting to repeat such observations at higher
resolution (the authors used an 8'' aperture), comparing e.g. sources in the central parsec with those in other areas close to the GC,
such as the Arches and Quintuplet clusters.\\
\begin{equation}
\label{EqWLR}
\beta_{\lambda_1 \lambda_2} = - \frac{ln(p_{\lambda_2} / p_{\lambda_1})}{ln(\lambda_2 / \lambda_1)}
\end{equation}
In more recent observations, \cite{hatano2013} studied the foreground polarization towards the GC (innermost 5.7 deg$^2$), observing in
the J-, H-, and Ks-bands. They found a significant flattening of the power-law wavelength dependence (see Eq. \ref{EqWLR}) towards longer
wavelengths, finding power-law indices of $\beta_{JH} = 2.08 \pm 0.02$ and $\beta_{HKs} = 1.76 \pm 0.01$. While polarimetric J-band observations
are not possible with NACO (the J-band filter is mounted in the same filter wheel as the Wollaston prism), we notice a similar flattening
when we compare the H- and Ks-band data of our last study \citep{buchholz2011} with the Ks- and Lp-band data presented here: we find
$\beta_{HKs} = 2.4 \pm 1.7$ and $\beta_{KsLp} = 0.40 \pm 0.67$. Despite the large errors, we are able to confirm that the observed
flattening of the wavelength dependence extends well into the Lp-band. The only value that is directly comparable ($\beta_{HKs}$) agrees with the
value found by \cite{hatano2013} within the uncertainties, but it is quite a bit higher. It has to be considered of course that our studies only
cover a much smaller region of the sky, and that local effects may play a significant role, as already stated in \cite{buchholz2011}. But the observed
trend supports the findings and conclusions of \cite{hatano2013} towards this innermost region of the GC as well, specifically that large aligned grains
must exert a significant influence.         
\subsection{Correlation with extinction}
\begin{figure*}[!t]
\centering
\includegraphics[width=\textwidth,angle=-90, scale=0.73]{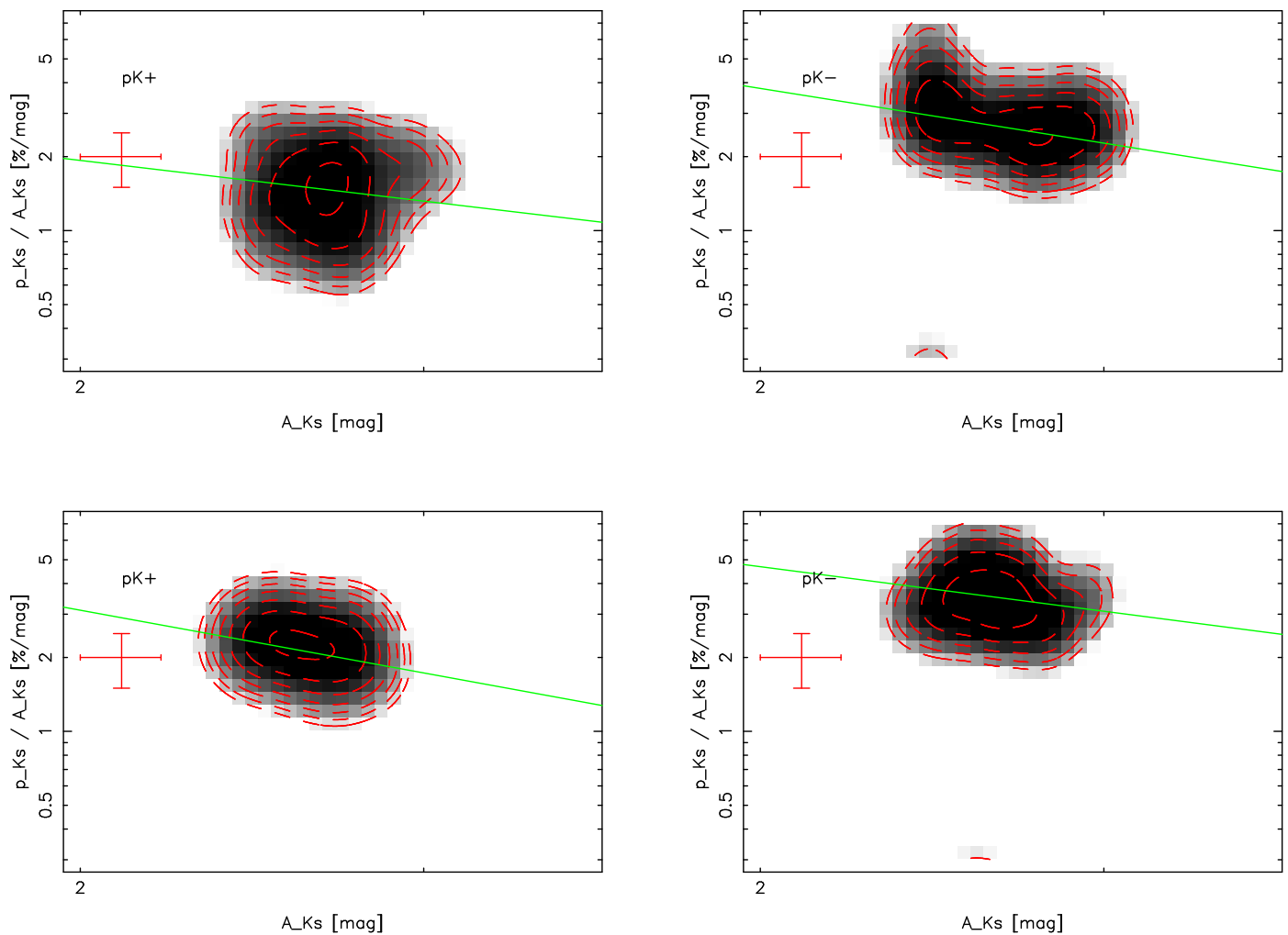}
\caption{\small Polarization efficiency in dataset K2 (Ks-band, upper frames) and dataset K3 (Ks-band, lower frames), compared
to Ks-band extinction, plotted as point density, with typical uncertainties represented on the left. pK$^+$ and pK$^-$ sources
shown separately in left respectively right frames for both bands. Green lines represent the fitted power-law relation. For
the polarization efficiency in dataset K1, see \cite{buchholz2011}, Fig. 16}
\label{FigPoleffK2K3}
\end{figure*}
\begin{figure*}[!t]
\centering
\includegraphics[width=\textwidth,angle=-90, scale=0.32]{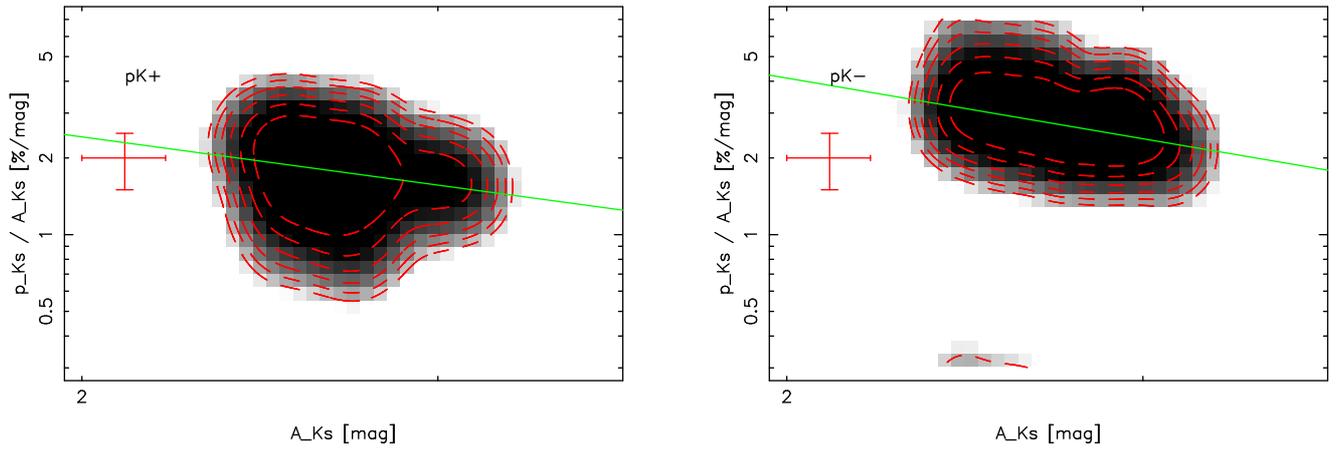}
\caption{\small Polarization efficiency of the sources in all Ks-band datasets (K1, K2, K3) compared to Ks-band extinction, plotted as
point density, with typical uncertainty represented on the left. pK$^+$ and pK$^-$ sources shown separately in left respectively right
frame. Green lines represent the fitted power-law relation.}
\label{FigPoleffKTotal}
\end{figure*}

We determined the extinction towards each source from the extinction map of the GC presented by \cite{schoedel2010b}. This was done in the
same way as in \cite{buchholz2011}.\\
For the sources found in the K2 and K3 datasets, a clear separation into two distinct groups of sources based on the Ks-band polarization
is not as evident as what was found in our previous study for the K1 dataset, but the polarization degrees still show a wide distribution
compared to the individual peaks found in \cite{buchholz2011} for the K1 data. In order to investigate if an offset in polarization
efficiency (as found in the previous work) can be found here as well, a separation at 6\% (K2) respectively 7.3\% (K3) was introduced
here (labeling the resulting sub-groups as pK$^+$ and pK$^-$) . These values were chosen based on Figs.\ref{FigPoldegsKnorth} and
Fig.7 in \cite{buchholz2011}. We excluded sources with less than 3\% polarization, since this low value indicates either a foreground source
or intrinsic polarization perpendicular to the foreground (e.g. in IRS~1W and 10W, see \S \ref{SectExtended}). In both of these cases,
no direct relation between the total polarization and the foreground extinction can be expected.\\ 
In all cases, the distributions can be fitted with a power law,
\begin{equation}
\frac{p_{\lambda}}{A_{\lambda}} \propto A_{Ks} ^{\beta},
\end{equation}
yielding the following power law indices:
\begin{eqnarray}
\beta_{K2,-} = -0.9 \pm 0.2 \nonumber \\
\beta_{K3,-} = -1.4 \pm 0.3 \nonumber \\
\beta_{K2,+} = -1.3 \pm 0.3 \nonumber \\
\beta_{K3,+} = -1.0 \pm 0.5. \nonumber
\end{eqnarray}
While slightly steeper than the power law found for dataset K1 previously, these new findings still agree with the results presented
in \cite{buchholz2011} and the conclusions drawn in that study. Considering the shape of the point density plots (see
Fig.\ref{FigPoleffK2K3}), the uncertainties given for the power law indices might even be underestimated. This effect is caused by
the larger scatter in the polarization parameters in these two datasets, which in turn results from the lower photometric quality.\\
Compared to the pK$^-$ values, a significant offset in polarization efficiency is detected for the pK$^+$ sources, while the
underlying power law appears to be very similar. Again, this reproduces the findings in our previous study for the smaller FOV of
dataset K1, and reinforces our conclusion that this might indicate a local contribution to the polarization, likely by dust in
the central parsec itself. It has to be noted, however, that the offsets found for the K2 and K3 sources are smaller compared to
the K1 sources, and this limits the confidence in this conclusion.\\
The distribution of the polarization efficiency combined for the three Ks-band datasets is shown in Fig.\ref{FigPoleffKTotal}. Fitting
a power law to the polarization efficiency leads to
\begin{eqnarray}   
\beta_{total,-} = -1.1 \pm 0.2 \nonumber \\
\beta_{total,+} = -1.3 \pm 0.2. \nonumber
\end{eqnarray}
These values agree with the results found for the individual datasets within the uncertainties. A similar offset between the higher and
lower polarized sources is also found here, but there is a considerable overlap between the sub-groups. This stems from the different
selection criteria applied to the different datasets.
\subsection{Bow-shock polarization}
\label{SectExtended}
\subsubsection{IRS~1W}
\label{SectIRS1W}
\begin{figure}[!t]
\centering
\includegraphics[width=\textwidth,angle=-90, scale=0.34]{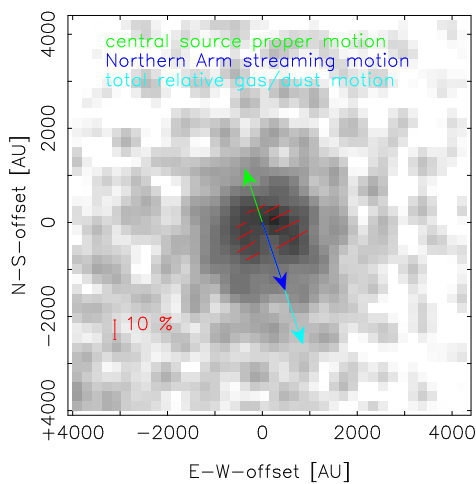}
\caption{\small Map of the intrinsic Lp-band polarization of the extended source IRS~1W. The arrows indicate the
proper motions of the central source, the motion of the Northern Arm material and the motion of both relative to each other.}
\label{FigIRS1W2dLp}
\end{figure}

In \cite{buchholz2011}, we determined the H- and Ks-band polarization of IRS~1W as
\begin{eqnarray}
p_H^{tot} &=& 5.2 \pm 0.5 \textnormal{ at } \theta_{H}^{tot} = -12^{\circ} \pm 5^{\circ} \nonumber\\
p_H^{int} &=& 6.9 \pm 0.5 \textnormal{ at } \theta_{H}^{int} = -73^{\circ} \pm 5^{\circ} \nonumber\\
p_{Ks}^{tot} &=& 1.8 \pm 0.5 \textnormal{ at } \theta_{Ks}^{tot} = -37^{\circ} \pm 5^{\circ} \nonumber\\
p_{Ks}^{int} &=& 7.8 \pm 0.5 \textnormal{ at } \theta_{Ks}^{int} = -75^{\circ} \pm 5^{\circ} \nonumber
\end{eqnarray}
The values found for the Lp-band confirm this trend: while a total polarization of (4.9 $\pm$ 0.5)\% at (-62 $\pm$ 5)$^{\circ}$ is measured,
applying a depolarization matrix with $p = 4 \%, \theta = 25^{\circ}$ (estimated based on the IRS~1 and IRS~16 point sources) within yields an intrinsic polarization of (8.9 $\pm$ 0.5)\% at
(-63 $\pm$ 5)$^{\circ}$. The polarization angle appears consistent over the three bands, which suggests that the same process is responsible,
while the increase of the intrinsic polarization degree towards longer wavelengths points to the higher influence of the extended dust component compared to that of the central source.\\ 
In order to investigate the polarization pattern in different regions of this extended source, the polarization was measured
in small apertures in the deconvolved images, and the foreground polarization was subtracted using a Mueller matrix as described
in \S \ref{SectForegroundremoval}. Fig.\ref{FigIRS1W2dLp} shows the polarization pattern in the Lp-band, and the similarity to both the H-
and the Ks-band pattern \citep[see][]{buchholz2011} is quite apparent, despite the much lower depth. The polarization increases
in the tails of the bow-shock, while the apex shows signs of depolarization.
\subsubsection{IRS~21}
\label{SectIRS21}
\begin{figure}[!t]
\centering
\includegraphics[width=\textwidth,angle=-90, scale=0.34]{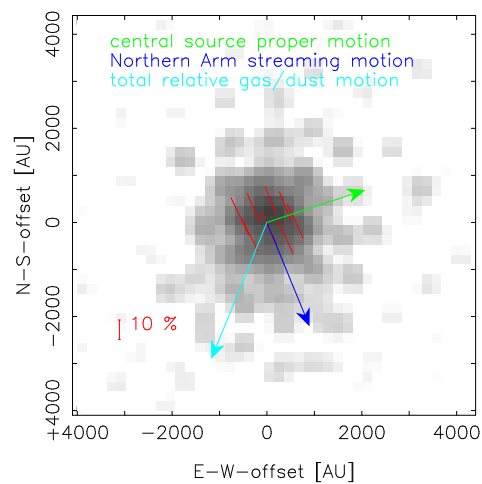}
\caption{\small Map of the intrinsic Lp-band polarization of the extended source IRS~21. The
arrows indicate the proper motions
of the central source, the motion of the Northern Arm material and the motion of both relative to each other.}
\label{FigIRS21Lp_2d}
\end{figure}

In \cite{buchholz2011}, we measured the Ks-band polarization of IRS~21 as
\begin{eqnarray}
p_{Ks}^{tot} &=& 9.1 \pm 0.5 \textnormal{ at } \theta_{Ks}^{tot} = 16^{\circ} \pm 5^{\circ} \nonumber\\
p_{Ks}^{int} &=& 6.1 \pm 0.5 \textnormal{ at } \theta_{Ks}^{int} = 5^{\circ} \pm 5^{\circ} \nonumber
\end{eqnarray}
In the Lp-band, IRS~21 shows a much stronger polarization compared to the Ks-band. The total polarization amounts to (19.0 $\pm$ 0.5)\% at
(24 $\pm$ 5)$^{\circ}$, and subtracting the foreground polarization (see \S \ref{SectIRS1W}, estimated based on the IRS~1 and IRS~16 point sources)
reveals an intrinsic polarization of (15.0 $\pm$ 0.5)\% at
(24 $\pm$ 5)$^{\circ}$.\\
By comparison, the polarization pattern in the Lp-band (see Fig.\ref{FigIRS21Lp_2d}) appears more uniform: polarization degrees
on the order of 15\% are detected in regions with significant flux, with very similar polarization angles of 20-25$^{\circ}$. As in the
Ks-band, no clear substructure (as it was found for IRS~1W) is apparent.
\subsubsection{IRS~10W}
\label{SectIRS10}
\begin{figure}[!t]
\centering
\includegraphics[width=\textwidth,angle=-90, scale=0.34]{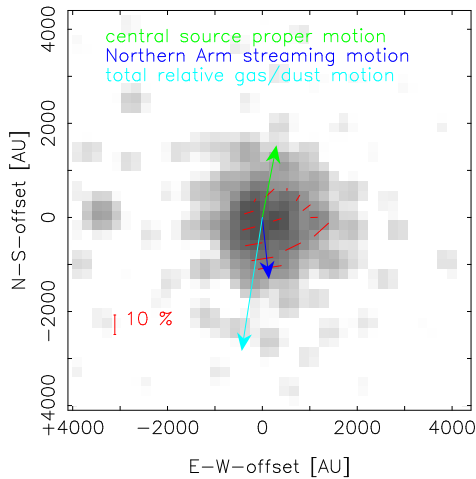}
\caption{\small Map of the intrinsic Lp-band polarization of the extended source IRS~10W. The
arrows indicate the proper motions
of the central source, the motion of the Northern Arm material and the motion of both relative to each other.}
\label{FigIRS10Lp_2d}
\end{figure}

IRS~10W is contained in the FOV of the 2011 Ks- and Lp-band data. No polarimetric H-band observations of this source are available.\\
The quality of this Ks-band dataset is insufficient for spatially resolved polarimetry, so only total values could be obtained: 
we measured the total polarization of IRS~10W as (2.1 $\pm$ 0.5)\% at (6 $\pm$ 5)$^{\circ}$ in the Ks-band. Applying a depolarization
matrix with the polarization of IRS~10E* used as the foreground value (this source is point-like and does not show signs of intrinsic
polarization, with (5.4 $\pm$ 0.5)\% at (28 $\pm$ 4)$^{\circ}$) yielded an intrinsic polarization of (4.2 $\pm$ 0.5)\% at
(-52 $\pm$ 5)$^{\circ}$. In the Lp-band, we found similar values: the total polarization is measured as (0.9 $\pm$ 0.5)\% at
(-61 $\pm$ 5)$^{\circ}$, which yields an intrinsic polarization of (5.6 $\pm$ 0.5)\% at (-67 $\pm$ 5)$^{\circ}$, assuming a
foreground polarization of (4.7 $\pm$ 0.5)\% at (22 $\pm$ 5)$^{\circ}$, which was again estimated based on IRS~10E*.\\
IRS~10W does not show the clear bow-shock morphology that was found in the case of IRS~1W (see
Fig.\ref{FigIRS10Lp_2d}). \cite{tanner2005} fitted this source with a bow-shock like shape, but the authors of that study themselves
cautioned that the angle of the best-fit solution did not agree with the observed proper motions of the source and the dynamics of
the Northern Arm. They instead suggested that the observed shape was produced by the additional influence of an outflow from IRS~10E*,
a highly variable source about 2'' to the south-east. The resolved Lp-band polarization (see Fig.\ref{FigIRS10Lp_2d})
is also not consistent with a bow-shock with the apex towards the east. The pattern found here would agree well with a bow-shock
oriented at about 30$^{\circ}$ East-of-North, which would be consistent with the source proper motions and the Northern Arm flow here.
The polarization pattern is less symmetric than that found for IRS~1W, and the uncertainties are larger, but these findings seem
to confirm the suggestion of \cite{tanner2005} that the bright region in the south-east is indeed not the apex of the bow-shock.\\
The total polarization measured in the Ks- and Lp-band supports this argument as well. An intrinsic polarization angle perpendicular
to the bow-shock angle would be expected, and therefore the observed angles of -52$^{\circ}$ respectively -67$^{\circ}$ would indicate
a bow-shock angle of about 20-40$^{\circ}$. 
\subsubsection{IRS~5}
\label{SectIRS5}
\begin{figure}[!t]
\centering
\includegraphics[width=\textwidth,angle=-90, scale=0.34]{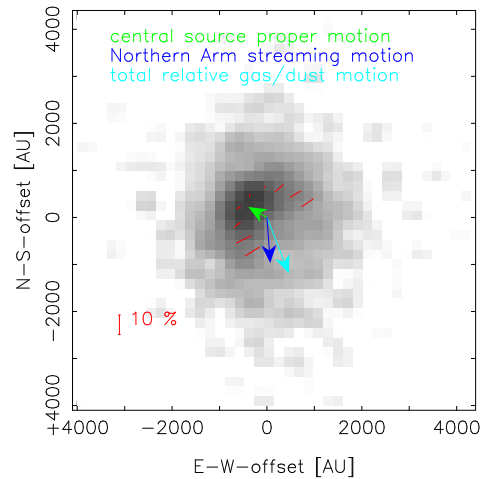}
\caption{\small Map of the intrinsic Lp-band polarization of the extended source IRS~5. The
arrows indicate the proper motions
of the central source, the motion of the Northern Arm material and the motion of both relative to each other.}
\label{FigIRS5Lp_2d}
\end{figure}

IRS~5 also shows a bow-shock shape in the Ks- and the Lp-bands \citep[as found by e.g.][]{tanner2005}, and the total polarization
measured in both bands at least partly agrees with the observed morphology: in the Ks-band, the source shows a total polarization
of (5.4 $\pm$ 0.5)\% at (24 $\pm$ 5)$^{\circ}$. Most sources in the vicinity seem to exhibit a stronger polarization (see
Fig.\ref{FigPolmapK}) on the order of 7.5\% at 24$^{\circ}$, based on sources within $\sim$5''. This yields an intrinsic polarization of (2.1 $\pm$ 0.5)\% at
(-62 $\pm$ 5)$^{\circ}$ for IRS~5. Note that this estimate depends critically on the accuracy of the foreground value.\\
Looking at the Lp-band, the source also shows a deviation in polarization from the sources around it: we measured a value of (1.8 $\pm$ 0.5)\%
at (14 $\pm$ 5)$^{\circ}$, and with a foreground polarization of 4\% at 25$^{\circ}$ this leads to an intrinsic value
of (2.5 $\pm$ 0.5)\% at (-57 $\pm$ 5)$^{\circ}$. The angle we find here coincides with the Ks-band angle and with the expectations
from the observed orientation of the bow-shock.\\
The resolved Lp-band polarization pattern (see Fig.\ref{FigIRS5Lp_2d}) resembles that of IRS~1W closely, with stronger polarization found
in the tails of the bow-shock, a decrease towards the apex and very uniform polarization angles. As for IRS~10W, the Ks-band
data quality does not allow spatially resolved polarimetry. The observed pattern in the Lp-band agrees very well
with the proper motions of the source and the streaming motion in the Northern Arm, as do the total intrinsic polarization parameters.
\subsubsection{Comparing the bright bow-shock sources}
In all four observed bright bow-shock sources, we find consistent polarization angles in all wavelength bands. In addition, there
appears to be a similar relation between the Ks/Lp-band polarization degrees for all but one source, IRS~21. This increases the
confidence in the respective foreground estimates (an incorrect foreground estimate can lead to a shift in the determined intrinsic
polarization degree and angle, and this would generally affect the Lp- and Ks-band results differently due to the different total values).\\
For IRS~1W, we find that the intrinsic Lp-band polarization degree exceeds the Ks-band value by a factor of 1.1. IRS~5 shows a similar
ratio (1.2), as does IRS~10W (1.3). By comparison, IRS~21 exhibits an increase from the Ks- to the Lp-band value by a factor of 2.5.
Both values again differ from the foreground value, where a ratio of 0.8 is known from previous observations \citep{jones1990} and
was again confirmed here. How can this behavior be explained?\\
\cite{buchholz2011} suggested emission from aligned grains as the dominant intrinsic polarization mechanism in the bow-shocks in the GC.
This model is supported by the resolved Lp-band polarimetry patterns presented here. But can the competing processes of scattering 
(single or multiple) and local dichroic extinction play a role as well? For a given grain alignment, dichroic extinction would produce
polarization perpendicular to emissive/single scattering polarization, while multiple scattering might simply reduce the observed
intrinsic polarization (although multiple scattering on efficiently aligned elongated grains may have the same result as single
scattering on these grains). The shape of IRS~21 suggests that the optical depth of the dust obscuring it is higher than what is found
towards the Northern Arm bow-shocks, but is it sufficient for a significant contribution of multiple scattering and/or local dichroic
extinction? While a detailed model of bow-shock polarization using measurements at more NIR/MIR wavelengths or even spectropolarimetry
as input parameters would be desirable to clarify the extent and relative importance of these processes, a rough estimate may already
be possible from the Ks- and Lp-band values alone.\\
The ratio $\frac{p_{Lp}}{p_{Ks}}$ is much higher for IRS~21 compared to the IRS~1W-type sources (1W, 5, 10W). \cite{moultaka2004} fitted the
Lp-band spectrum of IRS~21 with a 1200 K blackbody, which is hotter than the value found for IRS~1W in the same study ($\sim$900 K).
This alone would lead to a higher emission at shorter wavelengths, and if dust emission is dominant for both sources, why is the
polarization of IRS~1W and IRS~21 almost the same in the Ks-band, while the latter source has a much higher polarization in the
Lp-band? The answer might be that the optical depth respectively dust extinction also has to be considered: a higher optical depth
towards IRS~21 would cause the light from the central source to be mostly absorbed at shorter wavelengths (which produces polarization
perpendicular to the emissive polarization) or affected by multiple scattering (which depolarizes the light). Even if this contribution
stays the same in the Lp-band, the dust primarily emits at longer wavelengths, so the highly polarized dust emission becomes more and
more dominant towards the Lp-band. The emissive polarization in the Ks-band is in turn reduced by the absorption/multiple scattering
component. For the IRS~1W-type sources, these two processes would not play a significant role due to the much lower optical depth
towards the central source.\\
H-band polarimetry is only available for one bow-shock source, IRS~1W \citep[see][]{buchholz2011}. Combined with our new data, we find
a HKsLp-relation of the polarization degrees of 0.9:1:1.1 (normalized to $p_{Ks}$). For the sources only affected by foreground
polarization, we find a relation of 1.8:1:0.9 from comparing the peaks of the distributions of the polarization degrees of the
individual datasets (with the {\em caveat} that these do not cover the same FOV). Considering the different mechanisms responsible for the polarization
in the two cases and the different grain parameters (especially temperature, density and efficiency of alignment, also possibly the
size distribution), it is not surprising that the relation found for the bow-shocks does not match the LOS behavior.\\
The intrinsic polarization degrees of the Northern Arm bow-shocks seems to exhibit another trend: from IRS~1W over IRS~10W to IRS~5,
the intrinsic polarization degree decreases in the Ks- and the Lp-band. If we assume that the polarization is only caused by emission
in these three sources, this effect might indicate a decline in grain alignment (which may in turn be caused by a lower magnetic field strength)
towards the northern region of the Northern Arm. Viewed from the other side, this finding indicates that the field strength increases as the
initially wide Northern Arm \citep{paumard2004} is compressed as it is funneled towards the center. This behavior appears consistent, but
we must caution that this might be an overly simplified view since the polarization degree is not necessarily directly correlated with
the field strength \citep[see][]{aitken1998}. In addition, the bow shocks should produce local disturbances in the magnetic field, which complicates
any possible relation even more.\\  
In order to further clarify these relations and to search for possible substructures in polarization in IRS~21, additional Lp-band 
observations with higher depth would be required, as well as Ks-band observations of the Northern Arm region with sufficient data
quality for resolved polarimetry on IRS~5 and IRS~10W. Furthermore, deep Lp-band observations would allow direct measurements of the polarization
of the Northern Arm dust features, as a supplement to the \cite{aitken1998} MIR data. 
\subsubsection{IRS~5NE}
\label{SectIRS5NE}
\begin{figure*}[!t]
\centering
\includegraphics[width=\textwidth,angle=-90, scale=0.455]{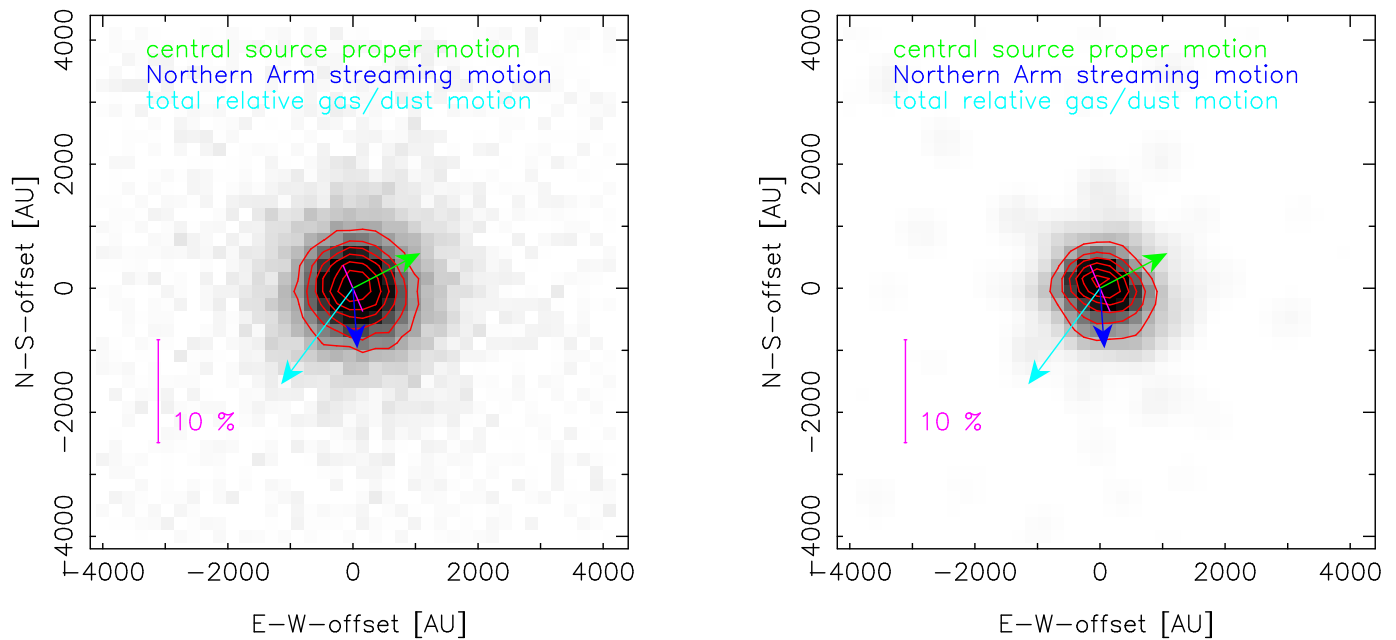}
\caption{\small Left: Lp-band image of the MIR excess source IRS~5E before LR
deconvolution. The arrows indicate the proper motions of the central source, the motion of the Northern Arm material and the
motion of both relative to each other. The magenta line indicates the polarization. Right: The same source after LR deconvolution.
The elongation of the source, the relative motion vector and the polarization all consistently indicate a bow-shock source.
Contours indicate flux levels, in  steps of 10\% of the maximum flux.}
\label{FigIRS5NELp}
\end{figure*}

In addition to the bright bow-shock sources observed in the GC, several other sources show a strong MIR excess. Among these,
IRS~5NE \citep[one of the sources east of IRS~5 that were examined in detail by][]{perger2008} shows a total Ks-band polarization
of (7.4 $\pm$ 0.5)\% at (13 $\pm$ 5)$^{\circ}$. The polarization parameters agree with the expected foreground polarization (see
\S \ref{SectIRS5}). In the Lp-band, there appears to be a significant polarization excess, with a total polarization of
(8.8 $\pm$ 0.5)\% at (24 $\pm$ 5)$^{\circ}$ compared to a value of 4\% at 25$^{\circ}$
in the vicinity. This leaves (4.8 $\pm$ 0.5)\% at (23 $\pm$ 5)$^{\circ}$ of intrinsic polarization. While \cite{perger2008} described
this source as compact with no apparent bow-shock morphology or other extended component, they suggest that the very red color
points to the influence of a dust envelope and that candidate sources include a dust-enshrouded low-luminosity AGB star or possibly
a young stellar object still in its dust shell. They argue that IRS~5NE might also be a low luminosity variant of the brighter bow-shock sources in the
central parsec.\\
With the higher resolution available here and after a Lucy-Richardson deconvolution,
IRS~5NE appears clearly extended in the Lp-band (although no apparent bow-shock shape can be observed, see Fig.\ref{FigIRS5NELp}).
The shape of the source would be marginally consistent with a
bow-shock considering its proper motions and the local streaming motion of the Northern Arm \citep[which seems to extend even further
eastward than IRS~5 and 10, see ][]{paumard2004}. The intrinsic polarization also suggests a dusty source, and the polarization angle
found here is almost perpendicular to the relative motion of the source through the surrounding material \citep{paumard2004,perger2008}
and the observed elongated feature. The source is too faint and the depth of the available data is too low for resolved polarimetry,
but the findings mentioned above already make a good case for IRS~5NE being a bow-shock source, similar to IRS~1W and IRS~5.\\
\subsubsection{IRS~2}
\label{SectIRS2}
\begin{figure*}[!t]
\centering
\includegraphics[width=\textwidth,angle=0, scale=1.0]{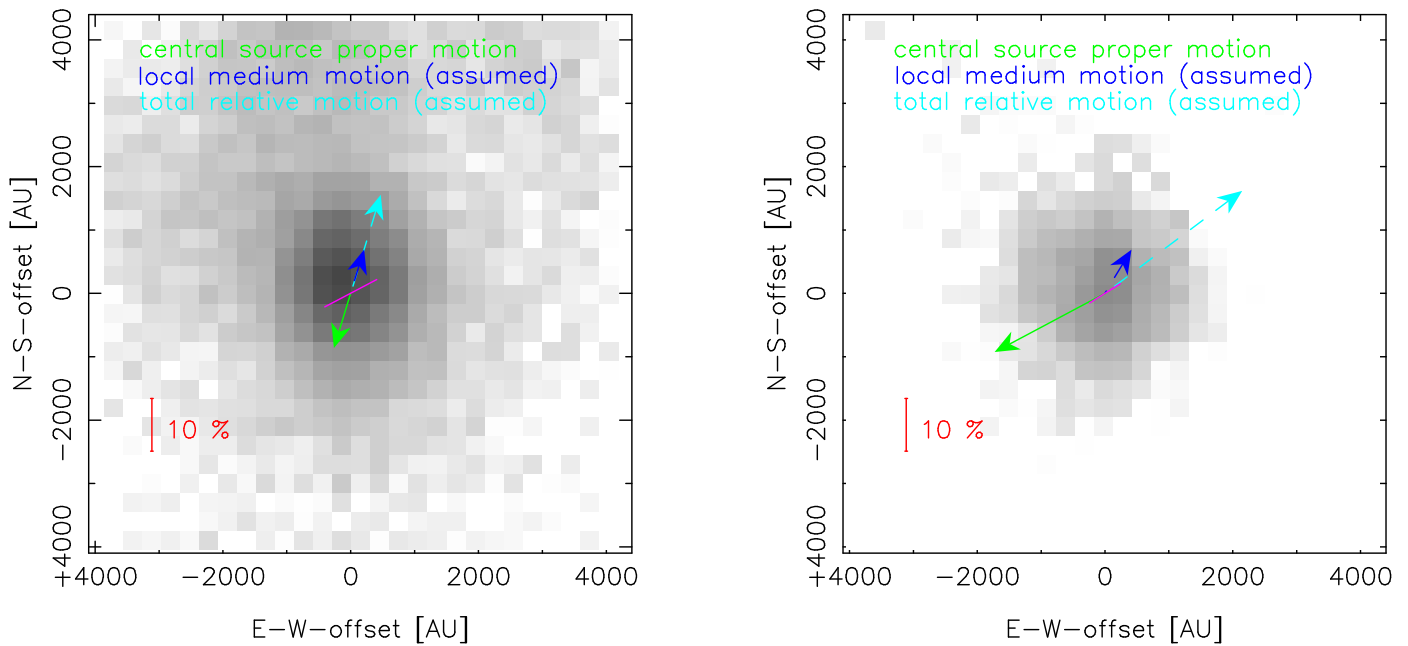}
\caption{\small Lp-band image of the MIR excess source IRS~2L (left) and IRS~2S (right) after LR
deconvolution. The arrows indicate the proper motions of the central source, the (assumed) motion of the local medium and the
(assumed) motion of both relative to each other. The magenta line indicates the polarization.}
\label{FigIRS2Lp}
\end{figure*}

The IRS~2 region, located south of IRS~13 \citep[see e.g. map by ][]{viehmann2005}, is made up of three main bright sources: the
extended sources IRS~2L in the north and IRS~2S in the south, and a more compact source between the two, here referred to as IRS~2C.
Polarimetric data is only available in the Lp-band for these sources. \\
A total polarization of (7.1 $\pm$ 0.5)\% at (-60 $\pm$ 5)$^{\circ}$ is found for IRS~2L, and with an
assumed foreground polarization of 4\% at 25$^{\circ}$ based on neighboring sources within $\sim$ 10'' , this source shows an intrinsic polarization of
(11.1 $\pm$ 0.5)\% at (-62 $\pm$ 5)$^{\circ}$. IRS~2S exhibits a weaker total polarization at a similar angle, with
(3.2 $\pm$ 0.5)\% at (-49 $\pm$ 5)$^{\circ}$. With the same foreground polarization, this leads to an intrinsic value of
(6.9 $\pm$ 0.5)\% at (-58 $\pm$ 5)$^{\circ}$. IRS~2C, located in projection directly between these two sources, shows
only the foreground value and no signs of intrinsic polarization. This can easily be explained if this source simply lies
in front of the dust feature that stretches from IRS~13 down to IRS~2, while the other two sources are embedded in this structure
and affected by it. This is supported by the shape of the sources in Lp-band images, where IRS~2C appears point-like and the other
two are visibly extended. The fact that both other sources show the same polarization angle but different polarization degrees may
indicate a different dust column density towards these sources, and indeed it seems as if the dust structure they are embedded in
becomes fainter towards the south \citep[see e.g. Fig.2 in][]{buchholz2011}.\\
No direct measurements of the projected velocities of the local medium are available in this region \citep[IRS~2 lies outside of the
region where velocities were mapped by ][]{paumard2004}. These sources lie between the regions for which \cite{zhao2009} determined
local gas velocities. But if the local magnetic field \citep[see ][]{aitken1998} follows the velocity
field as it is the case for the Northern Arm, this would indicate a local streaming motion with an angle of 20-30$^{\circ}$ to the
North-South axis. If these sources
are indeed embedded in a structure that is on an infalling orbit around the center (in a way similar to the Northern Arm), the local
streaming motion should be towards the north-west. Fig.\ref{FigIRS2Lp} shows an assumed motion of the local medium (of $\sim$120 km/s, dashed
blue line), and the resulting relative motion. If these were bow-shock sources, the polarization angle would be expected to be perpendicular
to the relative motion, and this is clearly not the case here: for IRS~2L, the observed angle differs by about 45$^{\circ}$ from the
expected value, while for IRS~2S, it is even perpendicular to what would be exptected for a bow-shock. In addition, the observed
angles differ about 30$^{\circ}$ from the MIR angles determined for the extended emission by \cite{aitken1998}. These findings indicate
that the observed polarization may be influenced significantly by the local extended dust distribution and the interaction of the
central sources with the medium. Without examining the spatial polarization patterns and comparing them over different wavelength bands,
it is not possible to determine whether these sources are bow-shocks or not.
\section{Conclusions}
\label{SectDiscussion}
We draw the following conclusions:\\
\begin{enumerate}
\item Our new Ks-band observations do not show clear localized offsets in polarization degree and angle, compared to our previous
study of the innermost 3''$\times$19'' \citep{buchholz2011}. These features may however be masked by the lower data quality, which
leads to a wider distribution of both parameters.
\item For the first time, Lp-band polarimetry was conducted with an 8m telescope, yielding polarization parameters for 84 sources
brighter than 11 mag in the central parsec. The results confirmed the findings by \cite{jones1990} of a deviation in the Lp-band
from the NIR wavelength dependency of the foreground polarization based on a much larger number of sources.   
\item The correlation between the spatially variable extinction towards the central parsec and the polarization efficiency,
$\frac{p_{\lambda}}{A_{\lambda}}$ that was determined in \cite{buchholz2011} has been confirmed here. Separating the newly observed 
sources based on their Ks-band polarization degree yields an offset in polarization efficiency similar to what was found in our
previous study. There may indeed be an additional local component of the polarization here as well, but it cannot be localized
as clearly as in the \cite{buchholz2011} data.
\item As already shown in e.g. \cite{buchholz2011}, intrinsic polarization takes place in several sources in the central parsec,
not only at longer wavelengths as shown
first by \cite{knacke1977}, but also in the H- and Ks-band \citep[e.g.][]{ott1999}. Using the M\"uller calculus, the intrinsic
component can be isolated for point sources and maps of extended features. The resulting total intrinsic
polarization angles for several known extended sources such as IRS~1W, IRS~5, IRS~10W and IRS~21 agree very well with what can be
expected from source morphology and relative motion of
gas/dust in the northern Arm and the sources themselves. Very similar intrinsic polarization degrees are measured for IRS~1W and
IRS~21, with (7.8 $\pm$ 0.5) \% at (-75 $\pm$ 5)$^{\circ}$ for IRS~1W and (6.1 $\pm$ 0.5) \% at (5 $\pm$ 5)$^{\circ}$ for IRS~21
(both Ks-band).
Contrary to the wavelength dependency seen in the foreground polarization, the H-band polarization degree is slightly lower
for IRS~1W compared to the Ks-band: (6.9 $\pm$ 0.5) \% at (-73 $\pm$ 5)$^{\circ}$. The fact that the central source contributes
a larger amount of flux in the H-band compared to the extended component, while the intrinsic polarization mostly stems from
the bow-shock, explains this discrepancy.\\
In the Lp-band, the intrinsic polarization degree of these two sources deviates considerably (while the angles are still very similar):
IRS~21 shows a much stronger Lp-band polarization of (15 $\pm$ 0.5)\% than IRS~1W with (8.9 $\pm$ 0.5)\%. This may be an effect of
a much higher dust density and therefore higher optical depth towards the former source. 
\item IRS~5 and IRS~10W show a relation between Ks- and Lp-band polarization similar to what was found for IRS~1W (at lower total
polarization degrees). It seems that the intrinsic polarization degree in both bands declines from IRS~1W to IRS~5, with IRS~10W
inbetween. This may point to a decrease in grain alignment (and therefore magnetic field strength?) from south to north along the
Northern Arm, but drawing such conclusions based on only three data-points is risky at best. The resolved patterns also resemble that
found for IRS~1W, and they also agree well with the observed proper motions and the streaming motion of the local medium, so it appears
that the same mechanism is at work there.  
\item Of the remaining MIR excess sources contained in the observed FOV, only IRS~5NE, IRS~2L and IRS~2S seem to show significant intrinsic
polarization (IRS~9, for instance, is apparently purely foreground-polarized). Considering the known proper motions of these sources
and the known (or in case of the IRS~2 sources, assumed) streaming velocities of the local medium, the polarization of IRS~5NE would agree
very well with a lower luminosity version of the bow-shocks observed in IRS~1W, 5 and 10W, while no clear conclusions can be drawn
for the IRS~2 sources. The latter may show an influence of local dust emission in addition to their own interaction with the medium.
\end{enumerate}
Several effects appear to contribute to the observed polarization in the central parsec of the GC, in addition to the foreground
polarization. While the latter is dominant in most sources in all three observed wavelength bands, large scale variations in foreground
polarization angle and degree, as well as intrinsic effects play a very important role and may be used to further investigate the
interstellar medium in the GC.\\
The recently achieved polarimetric calibration of NACO \citep{witzel2010} and the first successful polarimetric Lp-band observations
in nearly 30 years offer the possibility to use these tools to further enhance the knowledge of the stellar population and the ISM
in the vicinity of Sgr A*.  
\begin{acknowledgements}
We are grateful to all members of the NAOS/CONICA and the ESO PARANAL team. R. M. Buchholz acknowledges
support by Renia GmbH. R. Sch\"odel acknowledges support by the Ram\'on y Cajal programme of the Spanish Ministry of Economy and Competition, by
grants AYA2010-17631 and AYA2009-13036 of the Spanish Ministry of Economy and Competition, and by grant P08-TIC-4075 of the Junta
de Andaluc\'ia. Part of this work was supported by the COST Action MP0905: Black Holes in a violent Universe, the Deutsche Forschungsgemeinschaft
(DFG) via the Cologne Bonn Graduate School (BCGS), and via grant SFB 956, as well as by the Max Planck Society and the University of Cologne through
the International Max Planck Research School (IMPRS) for Astronomy and Astrophysics. We had fruitful discussions with members of the European Union
funded COST Action MP0905: Black Holes in a violent Universe and the COST Action MP1104: Polarization as a tool to study the Solar System and beyond.
We received funding from the European Union Seventh Framework Programme (FP7/2007-2013) under grant agreement No.312789. We would also like to
thank the anonymous referee for the helpful comments and suggestions.
\end{acknowledgements}


\begin{thebibliography}{}
\bibitem[Aitken et al.(1998)]{aitken1998} Aitken, D., Smith, C., Moore, T.,\& Roche, P., 1998, MNRAS 299, 743
\bibitem[Bailey et al.(1984)]{bailey1984} Bailey, J., Hough, J. H., \& Axon, D. J., 1984, MNRAS 208, 661
\bibitem[B\"oker(2010)]{boeker2010} B\"oker, T., 2010, IAU Symposium Proceedings, 266, 58-63
\bibitem[Buchholz et al.(2009)]{buchholz2009} Buchholz, R. M., Sch\"odel, R., \& Eckart, A. 2009, A\&A 499, 483
\bibitem[Buchholz et al.(2011)]{buchholz2011} Buchholz, R. M., Witzel, G., Sch\"odel, R., Eckart, A., Bremer, M., \& Muzic, K. 2011,
A\&A 534, A117
\bibitem[Capps \& Knacke(1976)]{capps1976} Capps, R. W., \& Knacke, R. F. 1976, ApJ 210, 76
\bibitem[Davis \& Greenstein(1951)]{davis_greenstein1951} Davis, L., \& Greenstein, J., 1951, ApJ 114, 206
\bibitem[Devillard (1997)]{devillard1997} Devillard, N., ''The eclipse software'', in The ESO messenger No 87 - March 1997
\bibitem[Eckart et al.(1995)]{eckart1995} Eckart, A., Genzel, R., Hofmann, R., Sams, B. J., \& Tacconi-Garman, L. E.
1995, ApJ 445, 23
\bibitem[Eckart et al.(2002)]{eckart2002} Eckart, A., Genzel, R., Ott, T., \& Sch\"odel, R. 2002, MNRAS 331, 917
\bibitem[Fritz et al.(2011)]{fritz2011} Fritz, T. K., Gillessen, S., Dodds-Eden, K., Lutz, D., Genzel, R., Raab, W.,
Ott., T., Pfuhl, O., Eisenhauer, F., \& Yusef-Zadeh, F. 2011, arXiv:1105.2282v2 
\bibitem[Ghez et al.(2003)]{ghez2003} Ghez, A. M., Duch\^ene, G., Matthews, K., Hornstein, S. D., Tanner, A., et al. 2003, ApJ
586, L127
\bibitem[Ghez et al.(2008)]{ghez2008} Ghez, A. M., Salim, S., Weinberg, N. N., Lu, J. R.,  Do, T., Dunn, J. K., Matthews, K.,
Morris, M., Yelda, S., Becklin, E. E., Kremenek, T., Milosavljevic, M., \& Naiman, J. 2008, ApJ 689, 1044
\bibitem[Gillessen et al.(2009)]{gillessen2009} Gillessen, S., Eisenhauer, F., Trippe, S., Alexander, T., Genzel, R., Martins, F.,
\& Ott, T. 2009, ApJ 692, 1075
\bibitem[Hatano et al.(2013)]{hatano2013} Hatano, H., Nishiyama, S., Kurita, M., Kanai, S., Nakajima, Y., Nagata, T., Tamura, M.,
Kandori, R., Kato, D., Sato, Y., Yoshikawa, T., Suenaga, T., \& Sato, S. 2013,  AJ 145, 105
\bibitem[Heiles(1987)]{heiles1987} Heiles, C. 1987, in Interstellar Processes, ed. D. Hollenbach and H. Thronson (Boston:Reidel), p.171
\bibitem[Jones(1989)]{jones1989} Jones, T. J. 1989, ApJ 346, 728
\bibitem[Jones(1990)]{jones1990} Jones, T. J. 1990, AJ, Vol. 99, Number 6, 1894
\bibitem[Knacke \& Capps(1977)]{knacke1977} Knacke, R. F., \& Capps, R. W. 1977, ApJ 216, 271
\bibitem[Kobayashi et al.(1980)]{kobayashi1980} Kobayashi, Y., Kawara, K., Kozasa, T., Sato, S., \& Okuda, H., 1980, Publ.
Astron. Soc. Japan 32, 291
\bibitem[Lazarian(2003)]{lazarian2003} Lazarian, A., 2003, Journal of Quantitative Spectroscopy \& Radiative Transfer,
v.79-80, p.881.
\bibitem[Lazarian et al.(2007)]{lazarian2007} Lazarian, A., \& Hoang, T., 2007, MNRAS, 378, 910
\bibitem[Lebofsky et al.(1982)]{lebofsky1982} Lebofsky, M. J., Rieke, G. H., Deshpande, M. R., \& Kemp, J. C., 1982, ApJ 263,
672
\bibitem[Lenzen et al.(2003)]{lenzen2003} Lenzen, R., Hartung, M., Brandner, W., et al. 2003, in SPIE Conf. Ser. 4841,
ed. M. Iye \& A. F. M. Moorwood, 944 
\bibitem[Martin et al.(1990)]{martin1990} Martin, P. G., \& Whittet, D. C. B., 1990, ApJ 357, 113
\bibitem[Mathis (1986)]{mathis1986} Mathis, J. S. 1986, ApJ 308, 281
\bibitem[Moultaka et al.(2004)]{moultaka2004} Moultaka, J., Eckart, A., Viehmann, T., Mouawad, N., Straubmeier, C., Ott, T.,
\& Sch\"odel, R. 2004, A\&A 425, 529
\bibitem[Nagata et al.(1994)]{nagata1994} Nagata, T., Kobayashi, N., \& Sato, S., 1994, ApJ 423, L113
\bibitem[Nishiyama et al.(2009)]{nishiyama2009} Nishiyama, S., Tamura,
  M., Hatano, H., Kanai, S., Kurita, M., Sato, S., et al., 2009, ApJ 690, 1648 
\bibitem[Nishiyama et al.(2010)]{nishiyama2010} Nishiyama, S., Hatano, H., Tamura, M., Matsunaga, N., Yoshikawa, Z., et al.
2010, ApJ 722, 23  
\bibitem[Ott et al.(1999)]{ott1999} Ott, T., Eckart, A., \& Genzel, R. 1999, ApJ 523, 248
\bibitem[Paumard et al.(2004)]{paumard2004} Paumard, T., Maillard, J.-P., \& Morris, M. 2004, A\&A 426, 81
\bibitem[Perger et al.(2008)]{perger2008} Perger, M., Moultaka, J., Eckart, A., Viehmann, T., Sch\"odel, R., \& Muzic, K.
2008, A\&A 478, 127
\bibitem[Purcell et al.(1971)]{purcell1971} Purcell, E.M., \& Spitzer, L., 1971, ApJ 167, 31
\bibitem[Reid et al.(2004)]{reid2004} Reid, M. J., \& Brunthaler, A. 2004, ApJ 616, 872
\bibitem[Rousset et al.(2003)]{rousset2003} Rousset, G., Lacombe, F., Puget, P., et al. 2003, in SPIE Conf. Ser. 4839,
ed. P. L. Wizinovich, \& D. Bonaccini, 140
\bibitem[Sch\"odel et al.(2002)]{schoedel2002} Sch\"odel, R., et al. 2002, Nature 419, 694  
\bibitem[Sch\"odel et al.(2003)]{schoedel2003} Sch\"odel, R., Ott, T.,
  Genzel, R., Eckart, A., Mouawad, N., \& Alexander, T. 2003, ApJ 596, 1015 
\bibitem[Sch\"odel (2010a)]{schoedel2010a} Sch\"odel, R. 2010a, A\&A 509, 58
\bibitem[Sch\"odel et al.(2010b)]{schoedel2010b} Sch\"odel, R., Najarro, F., Muzic, K., \& Eckart, A. 2010b, A\&A 511, 18
\bibitem[Sch\"odel(2010c)]{schoedel2010c} Sch\"odel, R. 2010c, The Milky Way Nuclear Star Cluster in Context, in: The Galactic
Center: a Window to the Nuclear Environment of Disk Galaxies. Proceedings of workshop at Shanghai, China, October 19-23, 2009,
ed. by M. Morris, Q. Wang, and F. Yuan. San Francisco: Astronomical Society of the Pacific, 2011, p.222, arXiv:1001.4238
\bibitem[Scoville et al.(2003)]{scoville2003} Scoville, N. Z., Stolovy, S. R., Rieke, M., Christopher, M. \& Yusef-Zadeh, F.
2003, ApJ, 594, 294 
\bibitem[Serkowski et al.(1975)]{serkowski1975} Serkowski, K., Mathewson, D. S., \& Ford, V. L., 1975, ApJ 196, 261
\bibitem[Tanner et al.(2002)]{tanner2002} Tanner, A., Ghez, A. M.,
  Morris, M., et al. 2002, ApJ 575, 860
\bibitem[Tanner et al.(2005)]{tanner2005} Tanner, A., Ghez, A. M., Morris, M., \& Christou, J. C. 2005, ApJ 624, 742
\bibitem[Viehmann et al.(2005)]{viehmann2005} Viehmann, T., Eckart, A., Sch\"odel, R., Moultaka, J., Straubmeier, C., \&
Pott, J.-U., A\&A 433, 117
\bibitem[Whitney \& Wolff (2002)]{whitney2002} Whitney, B. A., \& Wolff, M. J. 2002, ApJ 574, 205
\bibitem[Witzel et al.(2011)]{witzel2010} Witzel, G., Eckart, A., Buchholz, R. M., et al., 2011, A\&A 525, 130
\bibitem[Zhao et al.(2009)]{zhao2009} Zhao, J.-H., Morris, M. R., Goss, W. M., \& An, T. 2009, ApJ 699, 186 
\end{thebibliography}
\end{document}